# Counter-Propagating Core Assignment in Multi-Core Fiber Optical Networks to Reduce Inter-Core Crosstalk and Capacity Wastage

Fengxian Tang, Yonghu Yan, Limei Peng, *Member, IEEE*, Sanjay K. Bose, *Senior Member, IEEE*, Gangxiang Shen, *Senior Member, IEEE*

***Abstract*—Inter-core crosstalk is one of the most serious impairments for signal transmission in a multi-core fiber (MCF) optical network. On the other hand, because of wide deployment of data centers (DCs), we are seeing an increasing bidirectional traffic demand asymmetry, which leads to significant capacity wastage in designing and operating an optical transport network. To alleviate these effects, for an MCF optical network, we propose to assign fiber cores in an MCF in an asymmetric and counter-propagating manner. This can not only significantly reduce inter-core crosstalk between counter-propagating fiber cores but also flexibly assign different numbers of fiber cores in the opposite directions of a fiber link, thereby overcoming network capacity wastage due to the bidirectional traffic demand asymmetry. To evaluate the benefits of the proposed strategy, we consider the routing, spectrum, and core assignment (RSCA) problem for the MCF optical network. An integer linear programming (ILP) model and an auxiliary graph (AG) based heuristic algorithm are developed to optimize network spectrum resource utilization. Simulation studies show the effectiveness of the proposed core counter-propagation strategy, which can significantly outperform its counterpart, i.e., the co-propagation scheme, in terms of the total number of MCFs used and average inter-core crosstalk. In addition, the proposed RSCA heuristic algorithm is efficient to perform close to the ILP model, which can minimize the number of MCFs used and crosstalk between neighboring cores.**

## I. Introduction

The popularity of video-oriented applications such as Video on Demand (VoD) and Virtual/Augmented Reality (VR/AR) has led to the explosive increase of Internet traffic. Fiber-optic networks provide huge capacity for carrying Internet traffic. However, modern fiber-optic communication systems based on standard single mode single core fibers are near their transmission capacity limit [1]. To keep pace with the fast growth of Internet traffic, the capacity of the optical transport network should be increased and its transmission technology should be upgraded. Space division multiplexing (SDM) based optical transmission systems can significantly increase the transmission capacity of an optical transport network and therefore are being widely studied [2]. Among different types of SDM transmission systems, multi-core fiber (MCF) based transmission techniques are considered as one of the most promising and practical ones as their technology is more mature. We find that an MCF-based fiber-optic transmission system has been demonstrated to transmit nearly 1 Pb/s in a 32-core fiber [3]. However, one of the biggest challenges of MCF transmission is that inter-core crosstalk can severely degrade the quality of optical signals in two neighboring fiber cores, where the signals are transmitted in the same direction and wavelength. To reduce the inter-core crosstalk, many studies have been carried out for the assignment of spectrum and fiber cores when establishing lightpaths in an MCF optical network [4]. However, it seems that almost all the reported approaches have tried to reduce the inter-core crosstalk by sacrificing network capacity utilization. Specifically, in order to reduce the inter-core crosstalk, they avoid transmitting data on a wavelength in a core or using a lower-level modulation format on a wavelength when its neighboring core is carrying traffic on the same wavelength.

On the other hand, we also see that network traffic shows an increasing asymmetry. For example, video-oriented applications demonstrate a high bidirectional asymmetry. In these applications, traffic volume from a server to a user is typically much higher than that in the opposite direction. Secondly, network traffic also shows an increasing spatial asymmetry. Many data centers (DCs) that provide most contents in today's network are spatially dispersed in the network, which greatly increases the geographic asymmetry of traffic demand. The growth of network traffic asymmetry has led to a huge increase of capacity asymmetry in an optical transport network [5]. However, almost all the today's optical networks have been designed and operated based on the

Part of this work was presented in OFC 2018 [9]. This work was jointly supported by National Natural Science Foundation of China (NSFC) (61671313), and the Science and Technology Achievement Transformation Project of Jiangsu Province, PR China (BA2016123).

F. Tang and G. Shen are with the School of Electronic and Information Engineering, Soochow University, Suzhou, Jiangsu Province, P.R. China, 215006 (phone: 86-512-65221537; fax: 86-512-65221537; corresponding e-mail: Gangxiang Shen <shengx@suda.edu.cn>). Y. Yan is with Jiangsu Hengtong Fiber Technology Corporation, China. L. Peng is with the School of Computer Science and Engineering, Kyungpook National University, Daegu 41566, Korea. S. K. Bose is with the Dept. of EEE, IIT Guwahati, India.

assumption of bidirectionally symmetric traffic demands and capacities. This leads to significant wastage of network capacity and greatly reduces the network capacity utilization.

Reducing inter-core crosstalk and managing traffic demand bidirectional asymmetry are two challenging aspects for efficient design and operation of an MCF optical network. Therefore, how to efficiently design and operate such a network so as to significantly reduce inter-core crosstalk and capacity wastage due to bidirectional traffic demand asymmetry have become two key research issues. For the issue of inter-core crosstalk, the recent studies in [6][7] demonstrated that the counter-propagation of optical signals in the neighboring cores of an MCF can significantly reduce inter-core crosstalk. This motivated the key idea and novelty of this paper, i.e., to assign fiber cores in a counter-propagating way when assigning spectrum and fiber core resources in an MCF optical network. In addition, for the issue of bidirectional traffic demand asymmetry, our recent study in [8] found that significant savings can be achieved in both network design cost and capacity wastage due to traffic demand bidirectional asymmetry if an optical network is innovatively designed and operated in a unidirectional way and with decoupled transponders. This motivated us to consider the unidirectional design for an MCF optical network by assigning different numbers of cores in the opposite directions of a fiber link to tackle the traffic demand bidirectional asymmetry issue for better fiber and spectrum utilization.

The key contributions of this study are as follows. We propose to assign fiber cores in a counter-propagating manner to reduce inter-core crosstalk and to implement unidirectional design to tackle traffic demand bidirectional asymmetry for an MCF optical network. We also define a new parameter to estimate inter-core crosstalk according to the adjacency of two fiber cores. Based on our proposed counter-propagating assignment of fiber cores, we further consider the routing, spectrum, and core assignment (RSCA) problem based on the unidirectional design. The present paper significantly broadens and generalizes our preliminary study in [9]. Specifically, to fully explore the benefits of fiber core counter-propagation and unidirectional design for improving network capacity utilization and reducing inter-core crosstalk, we develop a novel ILP model for the RSCA problem as well as extend the RSCA heuristic algorithm in [9]. In the heuristic algorithm, different numbers of fiber cores are allocated in the two opposite directions of a fiber link according to the actual traffic demand in each direction. Moreover, the cores used to carry the traffic demand are selected in a crosstalk-aware manner based on a parameter used to estimate the inter-core crosstalk. Simulation results show that the strategy of counter-propagating core assignment in addition to unidirectional network design is effective in not only significantly reducing the inter-core crosstalk but also improving network capacity utilization. We find that our proposed heuristic algorithm is efficient and performs as well as the ILP model.

The rest of this paper is organized as follows. In Section II, we review related works on the inter-core crosstalk of MCFs and asymmetric network design. In Section III, we introduce the assignment of fiber cores in a counter-propagating manner for minimum inter-core crosstalk. We further present an ILP model and a heuristic algorithm for the above RSCA problem in Sections IV and V, respectively. Case studies and performance analyses are conducted in Section VI. Section VII concludes this paper.

## II. Related Works

In this part, we review existing works on inter-core crosstalk reduction in an MCF optical network as well as the design and operation of an optical network considering traffic demand bidirectional asymmetry.

### A. *Reducing Inter-core Crosstalk in MCF Optical Networks*

An SDM network with weakly-coupled MCFs is prone to signal impairment because of inter-core crosstalk [2]. It is important to address this issue when planning and operating an MCF optical network. Approaches proposed to cope with inter-core crosstalk can mainly be divided into two categories, i.e., *best-effort* and *strict constrained*, among which the best-effort category can further be divided into two sub-classes, i.e., *best-effort avoidance* and *best-effort core prioritization*.

The class of best-effort avoidance attempts to minimize inter-core crosstalk between neighboring cores when establishing a new lightpath. In [10], Zhao and Zhang proposed an efficient crosstalk-aware algorithm to minimize inter-core crosstalk between lightpaths. In [11], Fujii *et al.* proposed an on-demand routing and spectrum assignment (RSA) algorithm to mitigate inter-core crosstalk in an elastic MCF optical network. In [12], Muhammad *et al.* formulated the dimensioning problem of a programmable filterless SDM network with MCFs into an ILP model and related heuristic algorithms were developed.

The class of best-effort core prioritization achieves the same objective as the class of best-effort avoidance, but it additionally implements a dedicated core prioritization mechanism. Specifically, during spectrum and core allocation, cores are first ordered based on their priorities, where the priority of each core is determined by reducing dominant crosstalk based on a specific MCF structure. In [13], for an MCF structure, Toed *et al.* pre-determined the priority of each core and then assigned the cores according to their priorities. The approach can reduce the dominant inter-core crosstalk in an MCF optical network. In [14, 15], Fujii *et al.* proposed an on-demand spectrum and core assignment (SCA) method that can reduce both inter-core crosstalk and spectrum fragmentation in an elastic MCF optical network. In [16], Zhang *et al.* formulated an anycast routing, spectrum, and core assignment (ARSCA) problem into an ILP model, in which inter-core crosstalk is considered as an important objective to minimize. In [17], Toed *et al.* also proposed several routing, spectrum, core, and modulation assignment (RSCMA) algorithms. These algorithms have the advantages low computational complexities, flexibility in provisioning large-bandwidth services, spectral efficiency due to few spectrum fragmentations, and low inter-core crosstalk.

The strict constrained class relies on the estimation of inter-core crosstalk during lightpath establishment and network resource allocation. In particular, a lightpath can be provisioned for a user only when the inter-core crosstalk between both this new lightpath and other already established lightpaths is below a predefined threshold level. In [18], Muhammad *et al.* formulated the RSCA problem into an ILP model based on the estimation of inter-core crosstalk. Moreover, an efficient heuristic algorithm was developed to cope with the intractability of the ILP model for large networks. In [19], Muhammad *et al.* further proposed RSCMA algorithms which took into account inter-core crosstalk by adopting cross-layer optimization. In [20], Zhu *et al.* proposed a crosstalk-aware virtual optical network embedding (VONE) algorithm based on spectrum availability in an elastic MCF optical network. In [21], Li *et al.* formulated the routing, wavelength, and core allocation (RWCA) problem into an ILP model for an MCF optical network with multi-input-multi-output (MIMO) optical transmission technique. In [22], Li *et al.* investigated the shared backup path protection (SBPP) scheme for an MCF network taking into account additional routing constraints due to inter-core crosstalk and multiple concurrent failures. In [23], Li *et al.* estimated the capacity of an MCF optical network under limited digital signal processing (DSP) complexities and inter-core crosstalk. In [24], Perello *et al.* estimated the transmission reach of an optical signal over different types of MCFs with the consideration of amplified spontaneous emission (ASE) noise and inter-core crosstalk. In [25], Perello *et al.* proposed an optimal model and an efficient heuristic approach for the design of elastic MCF optical networks based on the worst case signal transmission reach. In [26], Dharmaweera *et al.* proposed a novel spectrum and core allocation scheme that considered both intra-core physical-layer impairments and inter-core crosstalk. In [27], Zhao *et al.* developed a new routing, spectrum, and core assignment algorithm that could establish mixed super-channels taking inter-core crosstalk considerations into account.

### B. Optical Network Design with Traffic Demand Bidirectional Asymmetry

For optical networks, there have been studies on efficient design of an optical network considering the asymmetry of bidirectional traffic demands. For the conventional dense wavelength division multiplexing (DWDM) optical network, Woodward *et al.* evaluated the impact of traffic demand asymmetry on a large IP backbone network and found that establishing unidirectional lightpaths and assigning wavelength resources according to the actual directional traffic demands can significantly reduce equipment costs [28]. In [29], Bathula and Zhang decoupled a transponder into a pair of independent transmitter and receiver, based on which they further explored the potential benefit of the proposed scheme in CAPEX savings when carrying asymmetric traffic demands. Morea *et al.* also evaluated the benefit of using elastic optoelectronic devices for carrying bidirectional asymmetric traffic demands in an IP over WDM optical network [30]. Other related works considering traffic demand bidirectional asymmetry or implementing unidirectional design for a DWDM network can be found in [31]-[35].

For an elastic optical network (EON), Ruiz and Velasco formulated the RSA problem into an ILP model for an EON that employed multicast trees to serve asymmetrical multicast demands and explored the benefit of cost savings by their proposed approaches [36]. In [37][38], Kim *et al.* formulated the RSA problem into an ILP model, in which flexible multi-flow optical transponders were employed to efficiently carry asymmetric traffic demands. In [39], Walkowiak and Klinkowski evaluated the impact of anycast and unicast traffic on transponder usage for both symmetric and asymmetric lightpath provisioning in an EON. Other related studies considering traffic demand bidirectional asymmetry or implementing unidirectional design for an EON can be found in [40]-[41].

### C. Summary and Novelty of This Work

Though different spectrum and fiber core assignment approaches have been proposed to reduce inter-core crosstalk in an MCF optical network, almost all the studies are based on the conventional assumption for optical transport network design, that is, a pair of opposite MCFs is set up on each fiber link and optical signals in each MCF are always transmitted in the same direction. Thus, designing an MCF optical network based on the counter-propagating core assignment in this study is novel, not considered before. Moreover, the benefit of employing this type of counter-propagation mode in reducing inter-core crosstalk has also not been evaluated before. It may be noted that most of the existing works considering traffic demand asymmetry have focused on the DWDM and elastic optical networks where a pair of standard single mode fibers is deployed on each link. However, few studies have focused on directly tackling the traffic demand asymmetry for an MCF optical network. In this study, the counter-propagation mode and multiple cores in each MCF create the freedom of assigning different numbers of cores in each signal propagation direction for efficient transmission of asymmetric traffic demands. This is innovatively different from the conventional asymmetric network design. Finally, the joint effort of reducing inter-core crosstalk and capacity wastage due to asymmetry of bidirectional traffic demands in an MCF optical network has not been examined before but will be explored in this paper.

## III. Core Counter-Propagation in an MCF Network with Bidirectional Asymmetric Traffic Demand

In the conventional optical network, there are typically two bidirectional fibers on each link. Though these two fibers carry signals in opposite directions, the signals in each fiber are co-propagating in nature. An MCF consists of multiple cores (e.g., 7 cores as shown in Fig. 1). Each core is spatially independent and they can transmit different optical signals. This allows cores in the same MCF to possibly transmit optical signals in opposite directions. Inter-core crosstalk is one of most serious transmission impairments in an MCF optical network. However, it has also been demonstrated that the counter-propagation of optical signals in the neighboring cores of an MCF can significantly reduce inter-core crosstalk [6], [7]. Thus, in this study we propose to assign fiber cores in a counter-propagating manner as shown in Fig. 1 so as to reduce inter-core crosstalk.

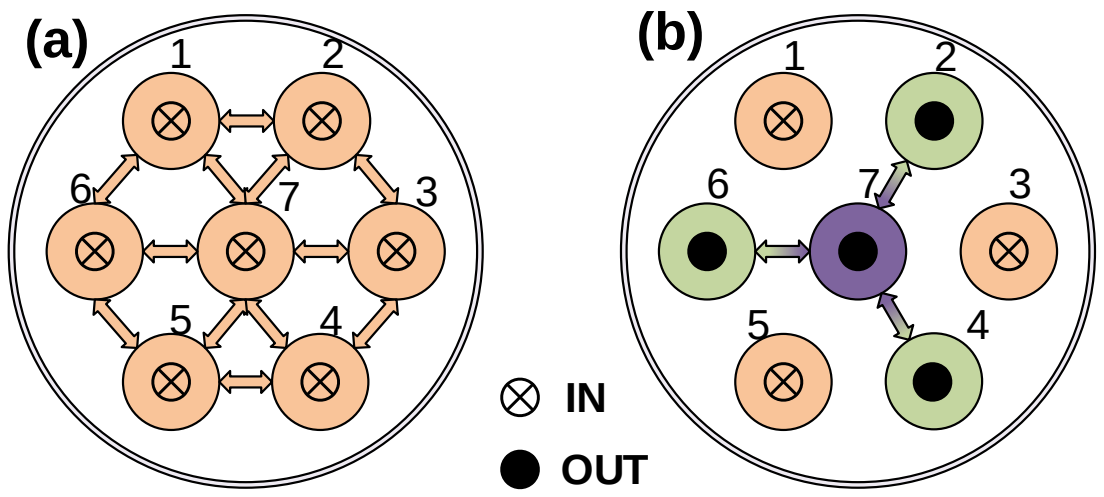

Fig. 1. Two core propagation modes. (a) Co-propagation mode; (b) Counter-propagation mode.

### A. Two Core Propagation Modes

Fig. 1 shows an example of a 7-core MCF, which includes two modes of optical signal propagation in its cores. Fig. 1(a) shows the *co-propagation mode*, under which all the fiber cores transmit optical signals in the same direction. This mode essentially follows the way that signals are transmitted in conventional optical networks, where each fiber link consists of a pair of fibers and all the signals in the same fiber transmit in the same direction while the signals in different fibers transmit in the opposite direction. In an MCF optical network, each MCF contains multiple fiber cores and each of them is spatially independent to transmit optical signals. This therefore allows the freedom of transmitting optical signals in the *counter-propagation mode* shown in Fig. 1(b). Specifically, we may assign different numbers of fiber cores in an MCF to transmit signals in the two directions. The benefits of this are twofold. First, the co-propagation mode in Fig. 1(a) suffers from severe inter-core crosstalk between optical signals in neighboring cores. In contrast, the counter-propagation mode in Fig. 1(b) can significantly suppress inter-core crosstalk by transmitting signals in the opposite directions as experimentally demonstrated in [6]. For a 7-core MCF, if the cores transmitting signals in the opposite directions are arranged in an interleaving manner as shown in Fig. 1(b), we can minimize inter-core crosstalk. Second, an increasing asymmetry of bidirectional traffic demand can lead to serious capacity wastage in an optical network designed and operated based on the symmetric traffic demand assumption. In an MCF optical network, there can be significantly higher capacity and core utilization in one link direction than the other. With the counter-propagation mode, we may assign different numbers of fiber cores in each MCF direction according to their actual traffic demands. This can largely avoid capacity wastage due to bidirectional traffic demand asymmetry.

### B. Inter-Core Crosstalk Factor

Depending on the location of each core in an MCF, the inter-core crosstalk between different core pairs is different. Fig. 2 shows an example of a 19-core MCF, in which there are three levels of crosstalk. The first level is L1, which exists between two cores that are directly neighboring, e.g., core 1 and core 2. The crosstalk between them is the strongest. The second level is L2, which exists between two cores where between them there is an intervening core, e.g., core 1 and core 14. Because the two cores are farther away compared to L1, this crosstalk is weaker than L1. Finally, the third level is L3, which exists between two cores where between them there are more than one intervening core, e.g., core 11 and core 17. It is reasonable that L3 has the lowest crosstalk since the two cores are the farthest from each other. To estimate the different levels of inter-core crosstalk, we define a weight factor $V_{i,j}$, where $i, j$ are the core indexes and $V_{i,j}$ represents the levels of inter-core crosstalk between the two cores. Because the first level of crosstalk is the strongest, its weight factor is the largest, while the other two levels of crosstalk are weaker, so their weight factors are smaller. In this study, we set the weight factor for the first level of crosstalk to be 100 and assign smaller weight factors for the second and third levels of crosstalk, which are 10 and 1, respectively.

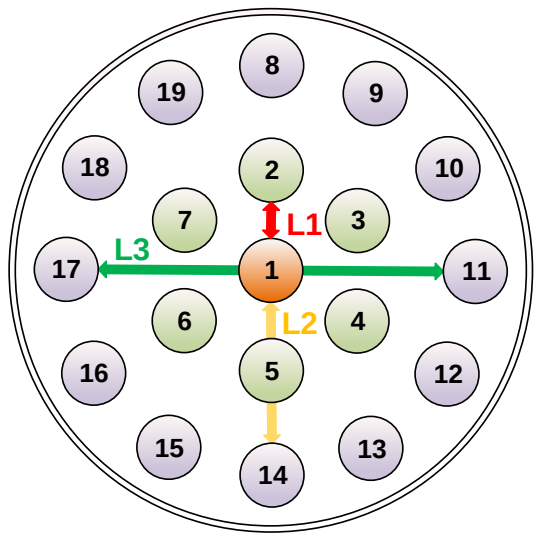

Fig. 2. Different levels of inter-core crosstalk.

It should be noted that inter-core crosstalk only exists when two lightpaths using the same spectrum and in different cores transmit in the same direction. Thus, to estimate the inter-core crosstalk, we first need to find the propagation directions of lightpaths and the spectra used by the lightpaths in two neighboring cores. If lightpaths are transmitted in the opposite directions, then no or little inter-core crosstalk would be suffered by the lightpaths according to [6]. In this study, we ignore inter-core crosstalk between lightpaths that are transmitted in the opposite directions. If the spectra of two lightpaths do not overlap, then also there would not be any inter-core crosstalk even when they transmit in the same direction.

Based on the adjacency level of fiber cores, lightpath transmission directions, and the amount of spectra overlapped by lightpaths, we further define a new term called *inter-core crosstalk factor* as

$$CF_{i,j} = V_{i,j} \cdot Z_{i,j} \cdot FS_{i,j} \quad (1)$$

Here $Z_{i,j}$ is a binary variable to denote whether cores $i$ and $j$ are carrying lightpaths in the same propagation direction. It equals 1 if they are in the same direction; 0, otherwise. $FS_{i,j}$ denotes the number of frequency slots (FSs) shared (or overlapped) by the lightpaths transmitted in cores $i$ and $j$. In the example of Fig. 3, we assume that there are lightpaths in cores 1, 2, and 5 transmitting in the same direction, and that the lightpaths in cores 8 and 14 transmit in the opposite direction. We assume that in core 1, a lightpath occupies FSs indexed from 1 to 3, and in core 2, two lightpaths occupy FSs indexed from 2 to 4 and from 18 to 23, respectively. Between cores 1 and 2, there are two FSs shared by their lightpaths, we denote this as $FS_{1,2} = 2$. Then the inter-core crosstalk factor can be calculated as $CF_{1,2} = 100 \times 1 \times 2 = 200$ since cores 1 and 2 have the first level of crosstalk. Similarly, we may calculate the inter-core crosstalk factor for cores 2 and 5 as $CF_{2,5} = 10 \times 6 = 90$, and for cores 8 and 14 as $CF_{8,14} = 1 \times 6 = 6$.

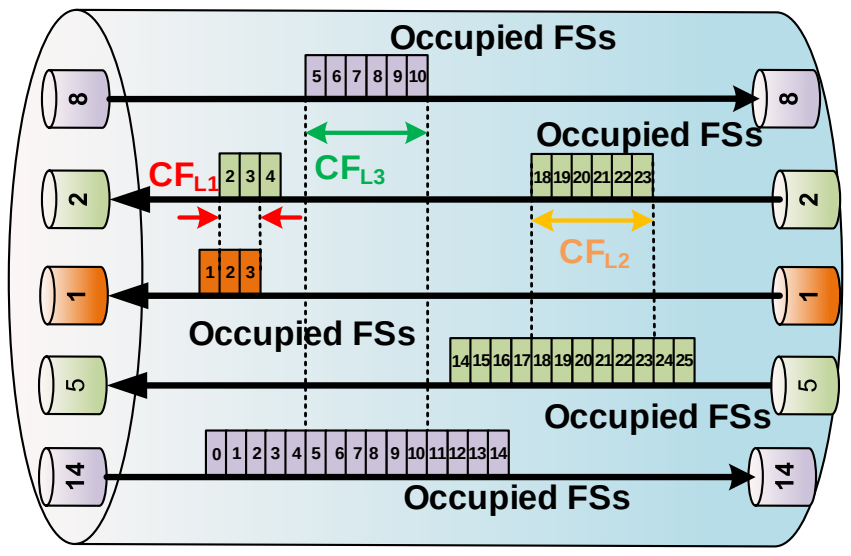

Fig. 3. Examples for the inter-core crosstalk factor.

## C. *MCF Optical Network with Counter-Propagating Cores*

We further introduce an MCF optical network that establishes lightpaths in the counter-propagating cores of MCFs, unlike what is done in a conventional bidirectional symmetric MCF network (see Fig. 4). Assume that there is a capacity demand of 4 FSs in the direction from B to C and a capacity demand of 8 FSs in the opposite direction. In a bidirectional symmetric design as shown in Fig. 4(a), a pair of opposite MCFs needs to be deployed between nodes B and C. Moreover, the larger capacity requirement between the two directions, i.e., 8 FSs, needs to be reserved for both directions even though the direction from B to C needs only 4 FSs. As a result, there are two fiber cores used and 4 FSs would be over-consumed (or wasted) in the direction from B to C. In contrast, under the core counter-propagation mode as shown in Fig. 4(b), only one MCF is needed between nodes B and C, where traffic demands in the two opposite directions are transmitted by a pair of counter-propagating cores. Moreover, the design based on counter-propagating cores can also efficiently assign spectrum and core resources based on the actual capacity requirements. Here, rather than the 8 FSs required by the bidirectional symmetric design, only 4 FSs need to be reserved for the direction from B to C. As this example shows, an MCF optical network based on counter-propagating cores can be more efficient in not only reducing the number of MCFs used, but also the spectra assigned to each core compared to what would have happened in a conventional symmetric design.

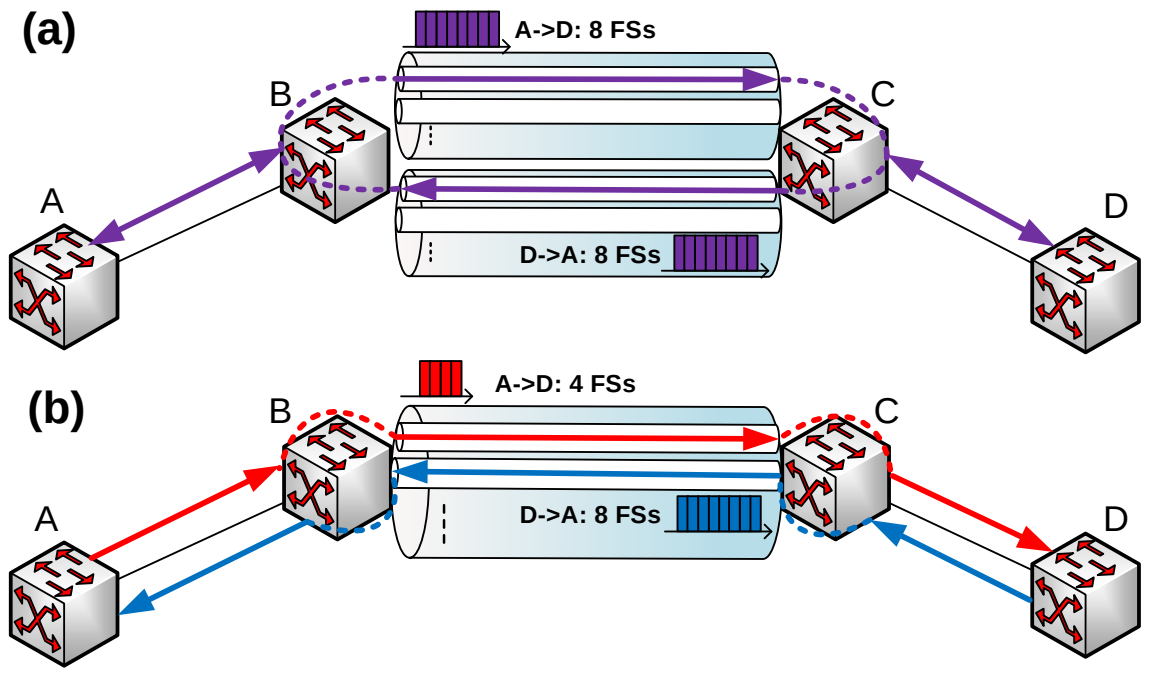

Fig. 4. Core assignments in two types of MCF networks. (a) Conventional bidirectional symmetric design; (b) Asymmetric design based counter-propagating cores.

# IV. Routing, Spectrum, and Core Assignment for an MCF Optical Network Based on Bidirectional Asymmetric Design

To evaluate the benefit of the proposed core counter-propagation mode in reducing inter-core crosstalk and capacity wastage due to traffic demand bidirectional asymmetry, we consider the routing, spectrum, and core assignment (RSCA) problem for the MCF optical network based on bidirectionally asymmetric design. In this section, we first present this RSCA problem, which is followed by an ILP model for the formulation of this problem.

## A. *Problem Statement*

The RSCA problem of an MCF optical network based on the core counter-propagation mode can be formally stated as follows.

*Given:*

1) A general network topology represented by a graph $G(\boldsymbol{N}, \boldsymbol{L})$, where $\boldsymbol{N}$ is the set of nodes and $\boldsymbol{L}$ is the set of fiber links connecting nodes in $\boldsymbol{N}$;
2) A set of lightpath demands given a priori. Each demand $r$ is represented by $r(S, D, FS)$, where $S$ and $D$ are the source and destination nodes of the demand, and $FS$ is the number of FSs required. The lightpath demand between a pair of nodes is directional and asymmetric.

*Constraints:*

1) Demand serving constraint: All the lightpath demands must be served.
2) Core constraint: There is a limited number of cores in each MCF.
3) Core capacity constraint: There is a limited number of FSs in each core.
4) Spectrum contiguity: The set of FSs allocated to a lightpath must be spectrally neighboring.
5) Spectrum continuity: The same set of contiguous FSs must be allocated on each link traversed by a lightpath.

The RCSA problem aims to minimize both the total number of MCFs used and the inter-core crosstalk between lightpaths by appropriately establishing lightpaths in the counter-propagating cores in an MCF optical network.

## B. *ILP Model*

For the RSCA problem, we develop an ILP model as follows.

**Sets:**

$\boldsymbol{L}$ The set of network links.

$\boldsymbol{C}$ The set of cores in each MCF.

$\boldsymbol{R}$ The set of node pairs. Under the unidirectional design, these node pairs are directional. That is, node pair A-B is not equivalent to node pair B-A. They are considered two different node pairs. For each node pair, we assume that there is a lightpath demand.

$\boldsymbol{P_r}$ The set of candidate routes used for establishing lightpaths between node pair $r \in \boldsymbol{R}$.

$\boldsymbol{B_r^p}$ The set of links traversed by route $p \in \boldsymbol{P_r}$ of node pair $r \in \boldsymbol{R}$.

**Parameters:**

$FS_r$ The number of FSs required by demand $r$.

$V_{i,j}$ A weight factor to represent the level of inter-core crosstalk between cores $i$ and $j$.

$\delta_{r1,p1}^{r2,p2}$ A binary parameter that equals 1 if route $p1$ between node pair $r1$ and route $p2$ between node pair $r2$ share common link(s); 0, otherwise.

$DL_{r,p}^{l}$ An integer parameter to denote the direction of link $l$ traversed by route $p$ between node pair $r$; it takes the value of 1 if the link is traversed from the *upstream* direction, 2 if traversed from the *downstream* direction. If link $l$ is not traversed by route $p$ between node pair $r$, it is 0. We use an example shown in Fig. 5(a) to explain the *upstream* and *downstream* directions. Specifically, for the node pair N1-N3 with a route N1-N4-N5-N3, link N1-N4 and N4-N5 traversed from a small node index to a larger node index, we define such a direction as *upstream* and set $DL = 1$; for link N5-N3 traversed from a large node index to a smaller node index, we define this as *downstream* and set $DL = 2$.

$F$ The maximum number of MCFs in each unidirectional network link.

$W$ The maximum number of FSs that each fiber core carries.

$M$ A large value.

$\varepsilon$ A small value.

$\alpha$ A weight factor.

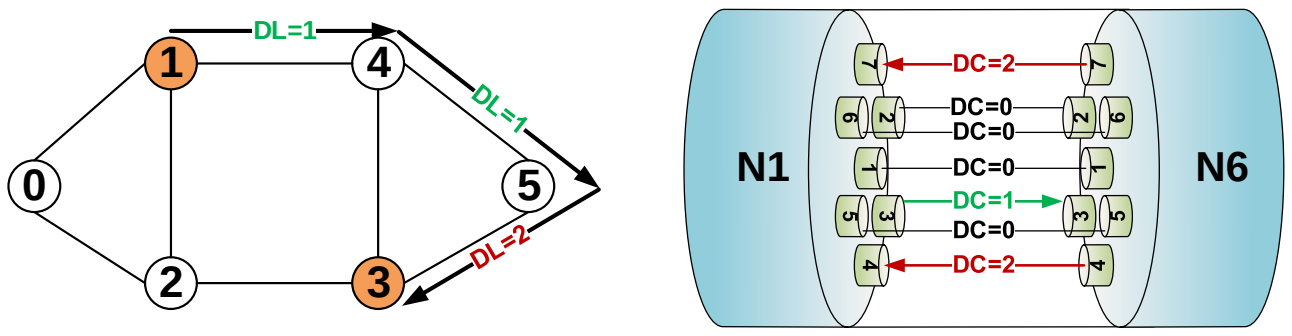

(a) DL (b) DC

Fig. 5. Definition of direction values.

**Variables:**

$f_{l,t}$ A binary variable that equals 1 if fiber $t$ of link $l$ is used; 0, otherwise.

$U_l^{i,t}$ A binary variable that equals 1 if core $i$ in fiber $t$ of link $l$ is used; 0, otherwise.

$S_{r,p}$ An integer variable denoting the starting FS index of the lightpath established on route $p$ between node pair $r$.

$E_{r,p}$ An integer variable denoting the ending FS index of the lightpath established on route $p$ between node pair $r$.

$\rho_{r1,p1}^{r2,p2}$ A binary variable that equals 1 if the starting FS index of the lightpath established on route $p1$ of node pair $r1$ is larger than that of the lightpath established on route $p2$ of node pair $r2$; 0, otherwise.

$X_p^r$ A binary variable that equals 1 if a lightpath is established on route $p$ between node pair $r$; 0, otherwise.

$O_{r,p}^{i,t,l}$ A binary variable that equals 1 if core $i$ in fiber $t$ of link $l$ is used for establishing a lightpath along route $p$ between node pair $r$; 0, otherwise.

$DC_l^{i,t}$ An integer variable denoting the direction value of core $i$ in fiber $t$ of link $l$, which equals 0 if not used, 1 if the direction is *upstream*, and 2 if the direction is *downstream*. In Fig 5(b), we use a 7-core MCF as an example to explain the *upstream* and *downstream* directions. In an MCF link N1-N6, if a core is used to transmit signals from a small node index to a larger node index, e.g., core 3, we define this direction as *upstream* and set $DC = 1$; if a core transmits signals from a large node index to a smaller node index, e.g., core 4 and core 7, we set $DC = 2$. Otherwise, if a core is not used, e.g., core 2 and core 6, we set $DC = 0$.

$Y_{t,l}^{i,j}$ A binary variable that equals 1 if cores $i$ and $j$ in fiber $t$ of link $l$ have the same direction value; 0, otherwise.

$\emptyset 1_{t,l}^{i,j}$ A binary variable that equals 1 if $DC_l^{j,t} \geq DC_l^{i,t}$; 0, otherwise.

$\emptyset 2_{t,l}^{i,j}$ A binary variable that equals 1 if $DC_l^{i,t} \geq DC_l^{j,t}$; 0, otherwise.

$Z_{t,l}^{i,j}$ A binary variable that equals 1 if both cores $i$ and $j$ in fiber $t$ of link $l$ are used and their propagation directions are the same; 0, otherwise. Note that this variable is different from $Y_{t,l}^{i,j}$ only for the condition that both cores are used.

$\beta_{r,p}^{k}$ A binary variable that equals 1 if $k \geq S_{r,p}$ where $k$ is an FS index; 0, otherwise.

$\gamma_{r,p}^{k}$ A binary variable that equals 1 if $k \leq E_{r,p}$ where $k$ is an FS index; 0, otherwise.

$\theta_{t,l}^{i,k}$ A binary variable that equals 1 if FS $k$ in core $i$ of fiber $t$ on link $l$ is used; 0, otherwise.

$A_{t,l}^{i,j,k}$ A binary variable that equals 1 if cores $i$ and $j$ in fiber $t$ of link $l$ are propagating signals in the same direction and both cores are using FS $k$ for lightpath establishment; 0, otherwise.

**Objective:**

$$\text{Minimize } \sum_{l\in \boldsymbol{L},1\leq t\leq F} f_{l,t} + \alpha \cdot \sum_{l\in \boldsymbol{L},1\leq t\leq F,i,j\in \boldsymbol{C},1\leq k\leq W:i\neq j} A_{t,l}^{i,j,k} \cdot V_{i,j}$$

Our objective is to minimize the total number of MCFs used and inter-core crosstalk between lightpaths in the whole network. Here $\alpha$ is a weight factor, which is set to be a small value such that the first objective has a higher priority. In this study, we set $\alpha = 0.01$.

**Subject to:**

**–Route selection**

$$\sum_{p\in \boldsymbol{P_r}} X_p^r = 1 \ \forall r \in \boldsymbol{R} \tag{2}$$

**–FS assignment**

$$E_{r,p} - S_{r,p} - FS_r + 1 \le M \cdot \left(1 - X_p^r\right) \ \forall r \in \boldsymbol{R}, p \in \boldsymbol{P_r} \tag{3}$$

$$E_{r,p} - S_{r,p} - FS_r + 1 \ge -M \cdot \left(1 - X_p^r\right) \ \forall r \in \boldsymbol{R}, p \in \boldsymbol{P_r} \tag{4}$$

$$E_{r,p} \le W \ \ \forall r \in \boldsymbol{R}, p \in \boldsymbol{P_r} \tag{5}$$

$$\rho_{r1,p1}^{r2,p2} + \rho_{r2,p2}^{r1,p1} = 1 \ \forall r1, r2 \in \boldsymbol{R}, p1 \in \boldsymbol{P_{r1}}, p2 \in \boldsymbol{P_{r2}}: r1 \ne r2 \tag{6}$$

$$E_{r2,p2} - S_{r1,p1} \le M \cdot \left(\rho_{r1,p1}^{r2,p2} + 1 - O_{r1,p1}^{l,i,t} + 1 - O_{r2,p2}^{l,i,t}\right) - 1 \ \forall r1, r2 \in \boldsymbol{R}, p1 \in \boldsymbol{P_{r1}}, p2 \in \boldsymbol{P_{r2}}, i \in \boldsymbol{C}, l \in \boldsymbol{B_{r1}^{p1}} \cap \boldsymbol{B_{r2}^{p2}}, 1 \le t \le F: r1 \ne r2 \tag{7}$$

**–Fiber core assignment**

$$\sum_{1\le t\le F, i\in \boldsymbol{C}} O_{r,p}^{i,t,l} - 1 \le M \cdot \left(1 - X_p^r\right) \ \forall r \in \boldsymbol{R}, p \in \boldsymbol{P_r}, l \in \boldsymbol{B_r^p} \tag{8}$$

$$\sum_{1\le t\le F, i\in \boldsymbol{C}} O_{r,p}^{i,t,l} - 1 \ge -M \cdot \left(1 - X_p^r\right) \ \forall r \in \boldsymbol{R}, p \in \boldsymbol{P_r}, l \in \boldsymbol{B_r^p} \tag{9}$$

$$U_l^{i,t} \ge O_{r,p}^{i,t,l} \ \ \forall r \in \boldsymbol{R}, p \in \boldsymbol{P_r}, l \in \boldsymbol{B_r^p}, i \in \boldsymbol{C}, 1 \le t \le F \tag{10}$$

$$f_{l,t} \ge U_l^{i,t} \ \ \forall l \in \boldsymbol{L}, i \in \boldsymbol{C}, 1 \le t \le F \tag{11}$$

**–Direction judgement**

$$DL_{r,p}^{l} \cdot O_{r,p}^{i,t,l} = DC_l^{i,t} \ \ \forall r \in \boldsymbol{R}, p \in \boldsymbol{P_r}, l \in \boldsymbol{B_r^p}, i \in \boldsymbol{C}, 1 \le t \le F \tag{12}$$

$$DC_l^{i,t} \le M \cdot U_l^{i,t} \ \ \forall l \in \boldsymbol{L}, i \in \boldsymbol{C}, 1 \le t \le F \tag{13}$$

$$1 - Y_{t,l}^{i,j} \le M \cdot \left(2 - \emptyset 1_{t,l}^{i,j} - \emptyset 2_{t,l}^{i,j}\right) \ \forall l \in \boldsymbol{L}, 1 \le t \le F, i, j \in \boldsymbol{C}: i \ne j \tag{14}$$

$$\varepsilon + DC_l^{i,t} - DC_l^{j,t} \le M \cdot \emptyset 1_{t,l_1}^{i,j} \ \ \forall l \in \boldsymbol{L}, i, j \in \boldsymbol{C}, 1 \le t \le F: i \ne j \tag{15}$$

$$\varepsilon - DC_l^{i,t} + DC_l^{j,t} \le M \cdot \emptyset 2_{t,l}^{i,j} \ \ \forall l \in \boldsymbol{L}, i, j \in \boldsymbol{C}, 1 \le t \le F: i \ne j \tag{16}$$

$$Z_{t,l}^{i,j} \ge Y_{t,l}^{i,j} + U_l^{i,t} + U_l^{j,t} - 2 \ \ \forall l \in \boldsymbol{L}, i, j \in \boldsymbol{C}, 1 \le t \le F: i \ne j \tag{17}$$

**–FS usage judgement**

$$k - S_{r,p} \le M \cdot \beta_{r,p}^{k} \ \ \forall r \in \boldsymbol{R}, p \in \boldsymbol{P_r}, 1 \le k \le W \tag{18}$$

$$E_{r,p} - k \le M \cdot \gamma_{r,p}^{k} \ \ \forall r \in \boldsymbol{R}, p \in \boldsymbol{P_r}, 1 \le k \le W \tag{19}$$

$$1 - \theta_{t,l}^{i,k} \le M \cdot \left(3 - \beta_{r,p}^{k} - \gamma_{r,p}^{k} - O_{r,p}^{i,t,l}\right) \ \forall r \in \boldsymbol{R}, p \in \boldsymbol{P_r}, l \in \boldsymbol{B_r^p}, i \in \boldsymbol{C}, 1 \le k \le W, 1 \le t \le F \tag{20}$$

**–Inter-core crosstalk judgement**

$$1 - A_{t,l}^{i,j,k} \le M \cdot \left(3 - \theta_{t,l}^{i,k} - \theta_{t,l}^{j,k} - Z_{t,l}^{i,j}\right) \ \forall r \in \boldsymbol{R}, p \in \boldsymbol{P_r}, l \in \boldsymbol{B_r^p}, i, j \in \boldsymbol{C}, 1 \le k \le W, 1 \le t \le F: i \ne j \tag{21}$$

**Route selection:** Constraint (2) ensures that only one route is selected for establishing the lightpath between a node pair.

**FS assignment:** Constraints (3) and (4) ensure the relationship between the starting and ending FS indexes when a specific route is selected for establishing a lightpath. Constraint (5) ensures that the ending FS index of any lightpath must be no greater than the maximum number of FSs that each fiber core carries. Constraints (6) and (7) ensure that the spectra of lightpaths are non-overlapping if the lightpaths are sharing a common fiber core in any fiber link. More specifically, the constraints ensure that if the starting FS index of lightpath A is larger than the starting FS index of lightpath B and they are sharing a common fiber core, then the starting FS index of the former must be also greater than the ending FS index of the latter.

**Fiber core assignment:** Constraints (8) and (9) mean that once a route is employed to establish a lightpath, a fiber core should be selected on each fiber link along the route to carry the lightpath. Constraint (10) means that once a core in an MCF is used for establishing a lightpath, then this core is considered used. Constraint (11) means that if any core in an MCF is used, then this fiber is used.

**Direction judgment:** Constraint (12) finds the propagation direction of a core. Constraint (13) means that if a core is not used, then the direction value must be 0. To take advantage of core counter-propagation for reducing inter-core crosstalk, we need to check the relative directions of neighboring cores. For this, constraints (14), (15), and (16) are derived to ensure the relationship that if $\left|DC_l^{i,m} - DC_l^{j,m}\right| = 0$, which means that cores $i$ and $j$ have the same direction value, then $Y_{m,l}^{i,j} = 1$. Constraint (17) derives the value for $Z_{t,l}^{i,j}$ based on $Y_{t,l}^{i,j}$ and $U_l^{i,t}$, which ensures $Z_{t,l}^{i,j} = 1$ if both cores $i$ and $j$ are used and they are in the same propagation direction.

**FS usage judgement:** Constraints (18), (19), and (20) jointly check whether FS $k$ is used in a fiber core of a fiber link.

**Inter-core crosstalk judgement:** Constraint (21) find whether an FS are used in both neighboring fiber cores when these two cores have the same propagation direction. Based on this, the objective term $\sum_{l\in \boldsymbol{L}, 1\le t\le F, i,j\in \boldsymbol{C}, 1\le k\le W: i\ne j} A_{t,l}^{i,j,k} \cdot V_{i,j}$ finds the total amount of inter-core crosstalk weighted by the *inter-core crosstalk factor* of the whole network.

### C. Complexity of ILP Model

The computational complexity of an ILP model is decided by the dominant numbers of variables and constraints. For the above ILP model, the computational complexity is analyzed as follows. The dominant number of variables are decided by the variables $O_{r,p}^{i,t,l}$ and $A_{t,l}^{i,j,k}$. For $O_{r,p}^{i,t,l}$, its number of variables is of order of $O(|\boldsymbol{C}| \cdot F \cdot |\boldsymbol{L}| \cdot |\boldsymbol{N}|^2 \cdot |\boldsymbol{P}|)$, where $|\boldsymbol{C}|$ is the number of cores in each fiber, $F$ is the number of fibers in each link, $|\boldsymbol{L}|$ is the number of network links, $|\boldsymbol{N}|$ is the number of network nodes, and $|\boldsymbol{P}|$ is the number of candidate routes between each node pair. For $A_{t,l}^{i,j,k}$, its number of variables is of order of $O(|\boldsymbol{C}|^2 \cdot W \cdot F \cdot |\boldsymbol{L}|)$, where $W$ is the number of FSs carried in each fiber core. Similarly, for the dominant number of constraints, we need to consider constraints (7) and (20). For (7), its dominant number of constraints is of the order of $O(|\boldsymbol{N}|^4 \cdot |\boldsymbol{P}|^2 \cdot |\boldsymbol{C}| \cdot |\boldsymbol{L}| \cdot F)$ and for (20), this number is of the order of $O(|\boldsymbol{N}|^2 \cdot |\boldsymbol{P}| \cdot F \cdot |\boldsymbol{L}| \cdot |\boldsymbol{C}| \cdot W)$.

## V. Heuristic Algorithm for RSCA Problem

The ILP model can find an optimal solution to the above RSCA problem. However, because the RCSA problem is NP-complete, for large or even reasonably sized networks, the ILP model cannot be solved to find an optimal solution within a reasonable time. Therefore, we also develop an efficient heuristic algorithm for the RSCA problem. To describe this algorithm, we first introduce the concept of spectrum window (SW) [42], which is used in the step of spectrum assignment so as to meet the constraints of spectrum contiguity and continuity. Based on this, we further describe the algorithm of crosstalk-aware counter-propagating core and spectrum assignment.

### A. *Concept of Spectrum Window (SW)*

For a lightpath request that needs a certain amount of capacity in units of FSs, we need to meet the constraints of spectrum *contiguity* and *continuity* when establishing this lightpath. Spectrum contiguity means that all the FSs of a lightpath must be spectrally neighboring. For this, we bring in the concept of spectrum window (SW) [42], which is made up of a set of continuous FSs. Fig. 6 shows an example of SWs on a fiber link where the size of each window is set to be 3 and there are a total of 10 SWs if the fiber link carries a total of 12 FSs. As shown in Fig. 6, each SW can always meet the constraint of spectrum contiguity. An SW is available only if all the contained FSs are free; otherwise, the SW is unavailable.

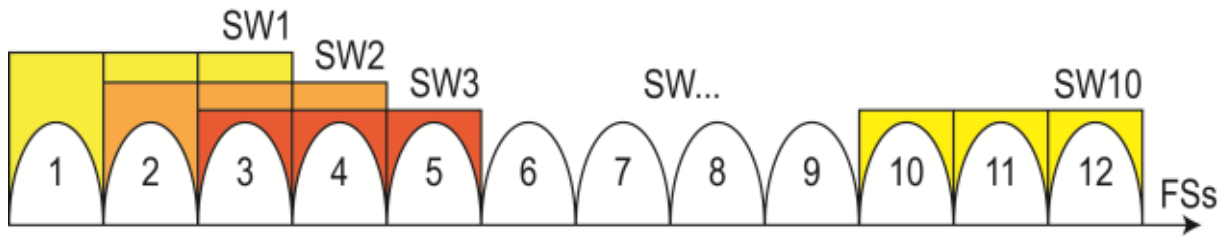

Fig. 6. Spectrum windows (SWs) in a fiber link.

Spectrum continuity is a type of spatial continuity, requiring all the fiber links traversed by a lightpath to use the same set of spectrally contiguous FSs. Given a route, to decide whether an SW is available along the route, we need to check if each of its traversed links is available of the corresponding SW. Only if all its links are available, is the SW considered available on the route [43]. Similar to a single fiber link, we can also create SWs for a route, as shown in Fig. 5. Based on this, we can then further select an available SW for lightpath establishment along the route.

### B. *Crosstalk-Aware Counter-Propagating Core Assignment*

To minimize inter-core crosstalk and avoid capacity wastage due to traffic demand bidirectional asymmetry in an MCF optical network, we propose an auxiliary graph (AG) based heuristic algorithm for the RCSA problem. The key idea of this algorithm is to assign neighboring fiber cores in an interleaving and counter-propagating manner. In addition, different numbers of fiber cores are assigned as per the actual capacity required in each direction. This avoids the capacity wastage that may happen because of the bidirectional asymmetry of the traffic demand. The key idea of this algorithm is to scan exhaustively all possible combinations of candidate routes and cores to select the one with the minimum number of newly deployed MCFs and the minimum inter-core crosstalk. The major steps of this algorithm are given as follows.

| | |
|---|---|
| Step 1: | Given a lightpath demand, use the K-shortest path algorithm to find a set of candidate routes between the node pair of the lightpath demand. |
| Step 2: | Try each $SW$ along each route $R$, denoted as $\langle R_i, SW_j \rangle$, and find the total number of new fibers required to be added, denoted as $l_{R,SW}$. |
| Step 3 | Find the set of combinations, denoted as $\{\langle R_i^*, SW_j^* \rangle\}$, which have the smallest $l_{R,SW}^{min}$, i.e., $\{\langle R_i^*, SW_j^* \rangle\} = \underset{\langle R_i, SW_j \rangle}{\mathrm{argmin}}\, l_{R_i, SW_j}$. |
| Step 4: | For each combination in $\{\langle R_i^*, SW_j^* \rangle\}$, create an auxiliary graph (AG), whose detail will be described later, and use Dijkstra's algorithm to find the least cost route and record the cost. |
| Step 5: | Based on different combination selection strategies (such as the least cost (LC) and first fit (FF) strategies), choose the corresponding cores and spectra. |

In Step 1, we find a set of candidate routes $R$ used for lightpath establishment based on the network topology of an MCF optical network.

In Step 2, for each route $R$, based on the current request that requires $f$ FSs, we scan each $f$-FS SW along the route to check whether each link can provide a free SW. Specifically, for each SW, we count the total number of links on the route $r$ that is not available of the SW. We denote this number as $l_{R,SW}$. If the route is available of the SW, then $l_{R,SW} = 0$, which means that there is no link that is not available of the SW. Otherwise, $l_{R,SW}$ would be greater than zero, which means that there is at least one link not available of the SW. We scan all the SWs for all the routes. After scanning all of them, we can generate a matrix as shown in Fig. 7 where the $y$-axis is the list of routes and the $x$-axis is the list of SWs on each of the routes. Each element in the matrix is $l_{R,SW}$.

In Step 3, based on the above matrix, we further find all the combinations of $R$ and $SW$ that have the smallest $l_{R,SW}^{min}$, i.e., $\{\langle R_i^*, SW_j^* \rangle\} = \underset{\langle R_i, SW_j \rangle}{\mathrm{argmin}}\, l_{R_i, SW_j}$. Note that there can be more than one combination of $R$ and $SW$ that has the smallest $l_{R,SW}^{min}$. For example, in Fig. 7, the combinations of $\langle R1, SW2 \rangle$ and $\langle R2, SWn \rangle$ both have the smallest $l_{R,SW}^{min}$.

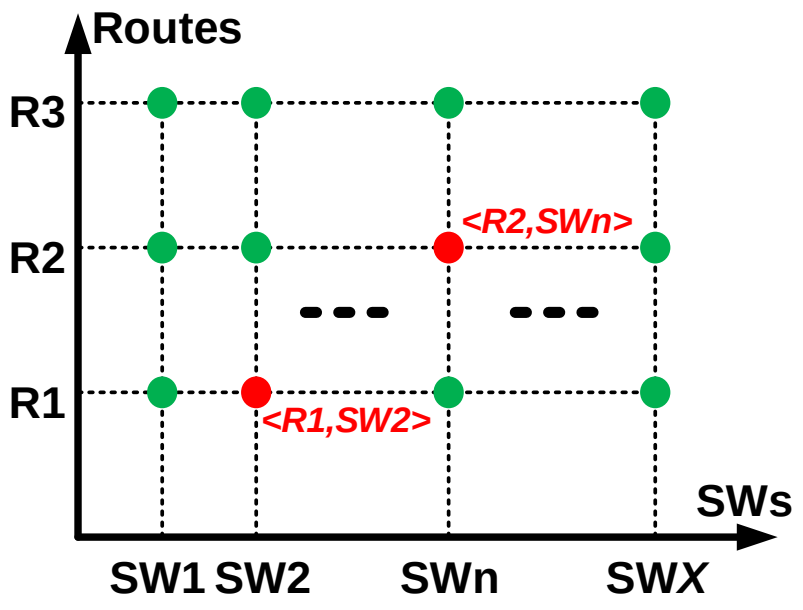

Fig. 7. Combinations of routes and SWs.

In Step 4, we create an AG for each combination in $\{\langle R_i^*, SW_j^*\rangle\}$ as shown in Fig. 8. In Fig. 8(a), an MCF optical network consists of two links. On each of the links, the usage and propagation direction of each fiber core are marked by an arrow. Also, the usage of spectrum resource in each fiber core is shown by the core. For example, core 2 in link Ns-N1 is propagating signals from Node 1 to Node s and its FSs with indexes from 1 to 3 are used. Based on this network, we create an AG as shown in Fig. 8(b).

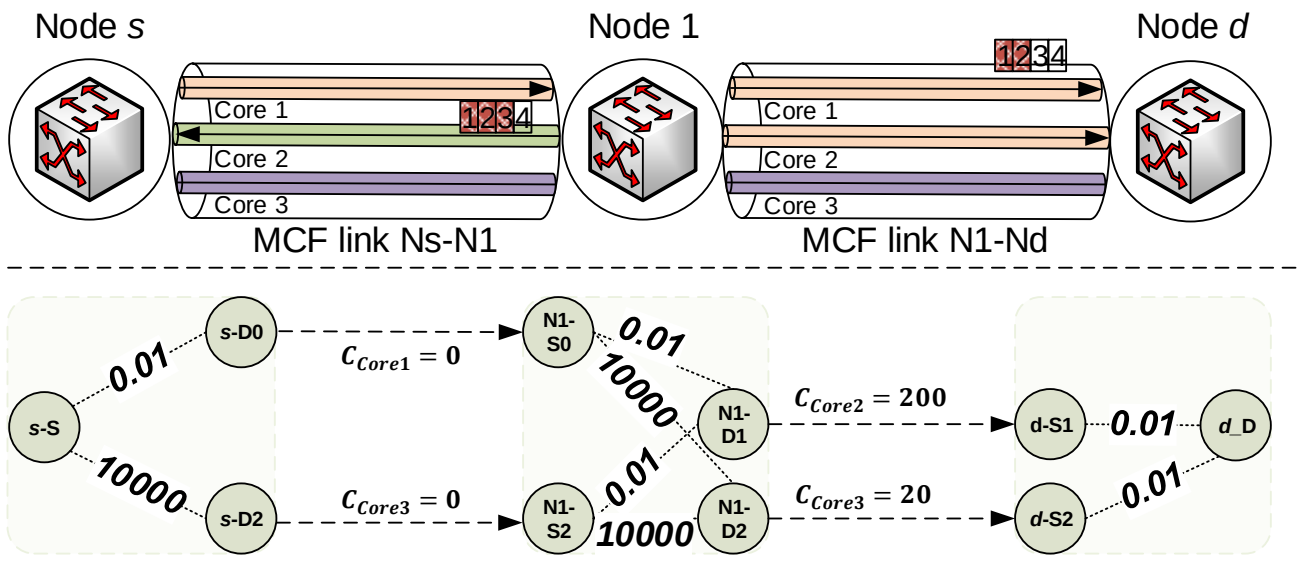

Fig. 8. Creating an auxiliary graph (AG).

Specifically, an MCF core that is not used yet or carries traffic in the direction from $s$ to $d$ but the considered SW (e.g., from FSs 1 to 4) is available is mapped to a unidirectional *auxiliary link* connecting two *auxiliary nodes*. For example, core 3 in link Ns-N1 is a core not used yet, so we create its corresponding auxiliary link connecting two auxiliary nodes, i.e., nodes s-D2 and N1-S2. Also, core 1 in link N1-Nd is used but it is available of the SW with FS indexes from 1 to 4. So we also create its corresponding auxiliary link connecting two auxiliary nodes, i.e., nodes N1-D1 and d-S1. For these core related links, we set their inter-core crosstalk factors as their costs. For example, in link Ns-N1, since core 2 carries traffic in the direction opposite to the direction from $s$ to $d$ and core 3 is not used, the inter-core crosstalk factor in the whole link is zero as according to (1), we set the costs for both auxiliary links as $C_{Core1} = 0$ and $C_{Core3} = 0$. Similarly, we can calculate the costs of the corresponding auxiliary links in link N1-Nd as $C_{core2} = 100 \times 1 \times 2 = 200$ and $C_{core3} = 10 \times 1 \times 2 = 20$ since core 1 has inter-core crosstalk with core 2 and core 3 in FSs indexed from 1 to 2.

Next, to inter-connect MCF cores via a switch node, we add auxiliary links to fully connect auxiliary nodes on both sides as shown. The cost of each auxiliary link is set as follows. If its destination virtual node corresponds to an unused MCF core (e.g., from N1-S0 to N1-D2 where core 3 is not used) then its cost is set to be large, e.g., $10^4$, to avoid using this unused core before using up spectrum resources on the other used cores. Otherwise, the cost is set to be small, e.g., 0.01. For nodes $s$ and $d$, auxiliary links are also added in a similar way.

Finally, based on the created AG, we further run the shortest path searching algorithm to find a path with the lowest cost and record the cost.

Note that in Step 4 there can be multiple combinations in $\{\langle R_i^*, SW_j^*\rangle\}$. Thus, in Step 5, we need select one combination from $\{\langle R_i^*, SW_j^*\rangle\}$ for lightpath establishment. For this, we consider two strategies, i.e., the first-fit (FF) and the least cost (LC) strategies. The FF strategy means to select the first combination in $\{\langle R_i^*, SW_j^*\rangle\}$ for lightpath establishment, while the LC strategy means to select a combination from $\{\langle R_i^*, SW_j^*\rangle\}$ that has the lowest cost of the route as found in Step 4.

### C. *Complexity Analysis*

The computational complexities of the heuristic algorithms are analyzed as follows. For the K-shortest path algorithm, computational complexity is of the order of $O(|\boldsymbol{R}| \cdot |\boldsymbol{N}|^2)$, where $|\boldsymbol{R}|$ is the number of candidate routes and $|\boldsymbol{N}|$ is the number of network nodes. In Steps 2 and 3, we need to find the combination of route and SW that has the smallest number of fibers newly added, which is of the order of $O(|\boldsymbol{R}| \cdot F \cdot |\boldsymbol{L}|)$, where the $F$ is the total number of FSs in each core and $|\boldsymbol{L}|$ is the number of network links. Then for each of selected route and SW combination, we construct an AG and run the shortest path algorithm. This step has the computational complexity of the order of $O(|\boldsymbol{N}|^2 \cdot |\boldsymbol{C}|^2)$, where $|\boldsymbol{C}|$ is the total number of cores and $|\boldsymbol{N}| \cdot |\boldsymbol{C}|$ is the total number of the nodes in a AG topology. Considering scanning all the combinations of route and SW selected in Step 3, the overall computational complexity in Step 4 is of the order of $O(|\boldsymbol{R}| \cdot F \cdot |\boldsymbol{N}|^2 \cdot |\boldsymbol{C}|^2)$.

## VI. Performance Analyses

We evaluated the effectiveness of the proposed counter-propagation strategy by running simulations based on three test networks, including (1) a six-node, eight-link (n6s8) network, (2) the 11-node, 26-link COST239 network, and (3) the 14-node, 21-link NSFNET network, as shown in Fig. 9. The distance of each link (in km) is shown next to the link. Both 7-core [8] and 19-core MCFs (see Fig. 2) are considered for this simulation study. The candidate routes used in the ILP model were obtained based on the link-disjoint $K$-shortest path algorithm. We employed the commercial AMPL/Gurobi software package (version 5.6.2) [44] to solve the ILP model, which was run on a 64-bit machine with 2.4-GHz CPU and 24-GB memory. The MIPGAP for solving the ILP model was set to be 0.01%.

To evaluate the efficiency of the proposed heuristic algorithm in comparison the ILP model, we ran simulations for these two schemes for the smallest test network, i.e., the n6s8 network, in which we assumed that there are a total of 200 unidirectional lightpath requests and each MCF has 7 cores with each core carrying 50 FSs. The capacity requirement of each unidirectional lightpath is uniformly distributed within the range of [1, 2$X$-1] FSs, where $X$ is the average number of FSs required. Note that the number of FSs assigned to each unidirectional lightpath can be derived from the actual capacity requirement between the corresponding node pair and the modulation format adopted according to the distance or signal quality of the lightpath.

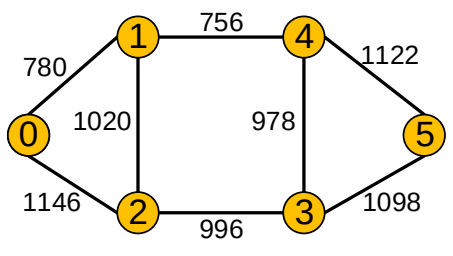

(a) 6-node, 8-link n6s8 network.

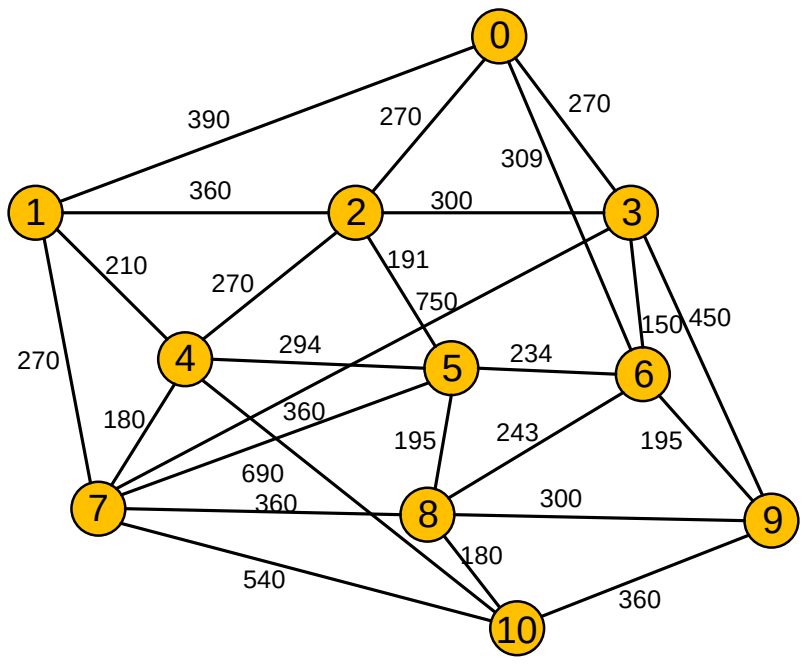

(b) 11-node, 26-link COST239 network.

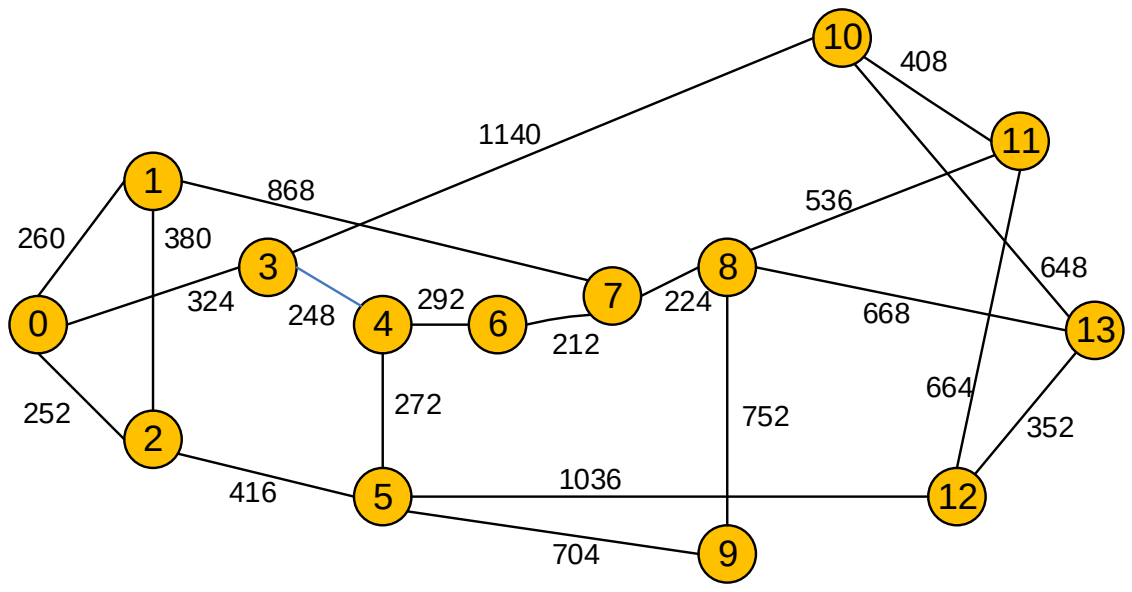

(c) 14-node, 21-link NSFNET network.

Fig. 9. Test networks.

For the heuristic algorithm, we also ran simulations for the other two larger test networks, i.e., the COST239 and NSFNET networks. For these, each MCF is assumed to have 7 cores with each core carrying 320 FSs. In addition, 1000 unidirectional lightpath requests were simulated. The capacity requirement of each unidirectional lightpath is uniformly distributed within the range of [5, 2$X$-5] FSs. In addition, considering the performance dependence on the order of lightpath demands provisioned, we shuffled an initial lightpath demand list 1000 times to form a set of shuffled demand sequences, and for each of the sequences, we ran the heuristic algorithm to find an RSCA solution and then selected the one with the best performance as our final solution.

For all the simulation cases, to account for the traffic demand bidirectional asymmetry, we assign different capacities to two unidirectional lightpaths between the same node pair. Specifically, a larger bandwidth is always assigned to a unidirectional lightpath whose source node index is larger than that of the destination node.

## A. *Number of MCFs Used and Inter-Core Crosstalk*

In this section, we compare the performance of the different schemes in terms of the number of MCFs used and the average inter-core crosstalk per FS of each channel, calculated as $\overline{CF} = \sum_{l\in \boldsymbol{L}, i,j\in \boldsymbol{C}, 1\le t\le F, 1\le k\le W: i\neq j} A_{t,l}^{i,j,k} \cdot V_{i,j} \,/\, \sum_{d\in \boldsymbol{D}} FS_d$, $\boldsymbol{L}$ is the set of network links, $\boldsymbol{C}$ is the set of cores, $\boldsymbol{D}$ is the set of unidirectional optical channels established, $F$ is the total number of fibers, and $W$ is the number of FSs in each core. $\sum_{l\in \boldsymbol{L}, 1\le t\le F, i,j\in \boldsymbol{C}, 1\le k\le W: i\neq j} A_{t,l}^{i,j,k} \cdot V_{i,j}$ finds the total amount of inter-core crosstalk weighted by the *inter-core crosstalk factor* in the whole network, and $FS_d$ is the number of FSs required by optical channel $d$.

Fig. 10 compares the total number of MCFs used and average inter-core crosstalk with an increasing capacity requirement for the small network n6s8. The legends "Counter" and "Co" represent the design cases of core counter-propagation and co-propagation, respectively. Here co-propagation corresponds to the conventional symmetric network design, in which a pair of MCFs is always set up per link and the two MCFs are transmmiting signals in the opposite directions. The legends of "FF" and "LC" correspond to the core and spectrum selection strategies in the AG-based algorithm, respectively. "ILP" corresponds to the ILP model.

We can see that with an increasing capacity requirement, all the schemes show requirements for more MCFs, which is reasonable since the increasing demand for more capacity requires more fibers. Comparing the cases of core counter-propagation and co-propagation, we can see that the counter-propagation strategy is effective in significantly reducing the number of MCFs required, by up to 54% and 53% under the FF and LC strategies, respectively. In addition, we see that the LC strategy outperforms the FF strategy by more than 10% because the former selects a combiantion of route and SW with the fewest newly added MCFs and the least inter-core crosstalk. We also compare the performance of the heuristic algorithm and the ILP model. The result shows that the counter-propagation scheme with the LC strategy can achieve performance very close to the ILP model. This shows the effectiveness of the proposed heuristic algorithm in reducing the number of MCFs used.

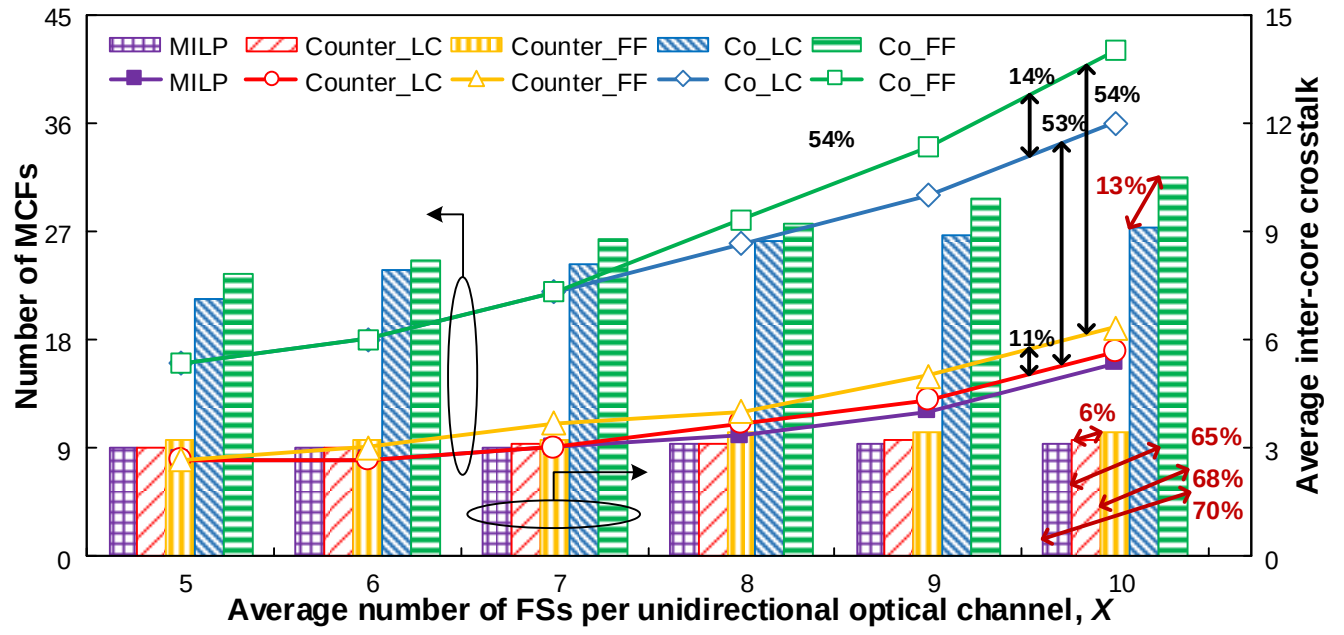

Fig. 10. Performance comparison in terms of number of MCFs used and inter-core crosstalk.

Similar observations can be made for the performance in terms of inter-core crosstalk. We can see that the use of counter-propagation significantly reduces the inter-core crosstalk by up to 68% and 65% compared to its counterpart, i.e., the case of co-propagation under the FF and LC strategies respectively. Also, the LC strategy achieves better performance than the FF strategy with 6% or more lower inter-core crosstalk. This is because the LC strategy scans all possible spectrum and core assignment scenarios to choose the one with the lowest crosstalk. Moreover, the counter-propagation scheme with the LC strategy performs very close to that of the ILP model, which however reduces up to 70% inter-core crosstalk compared to the worst case, i.e., co-propagation with the FF strategy. This therefore confirms the efficiency of the proposed heuristic algorithm in reducing inter-core crosstalk.

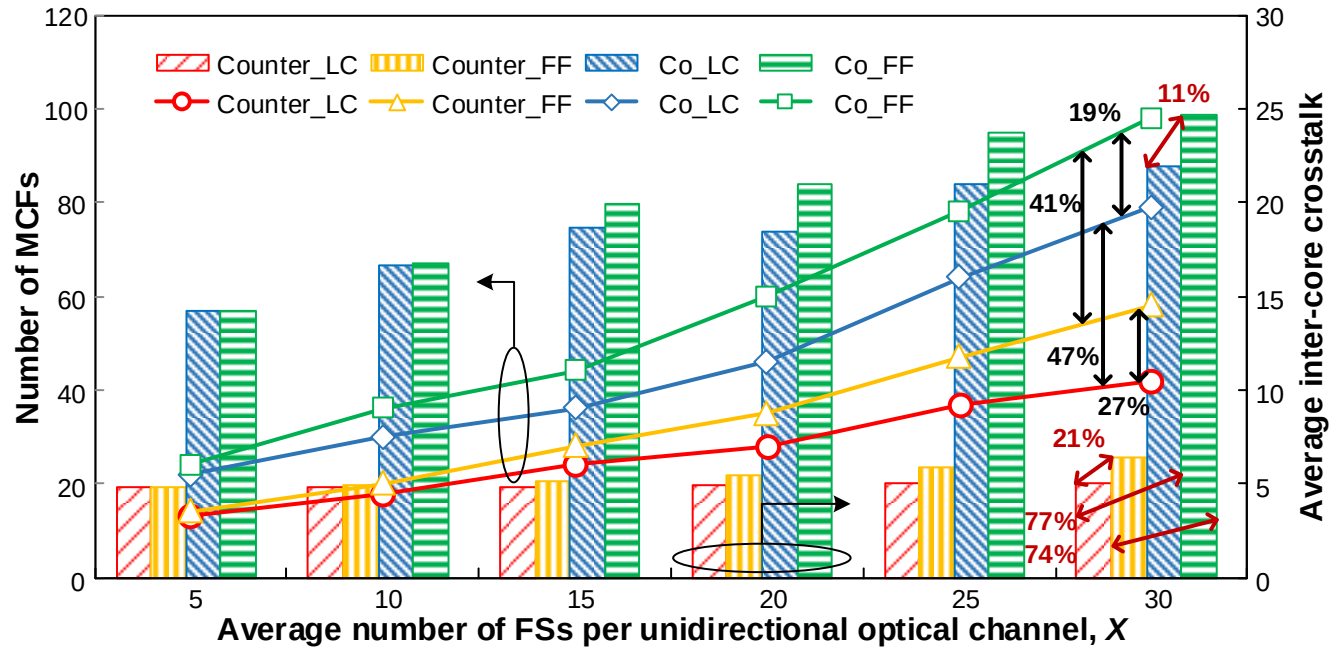

(a) 7-core MCF (NSFNET)

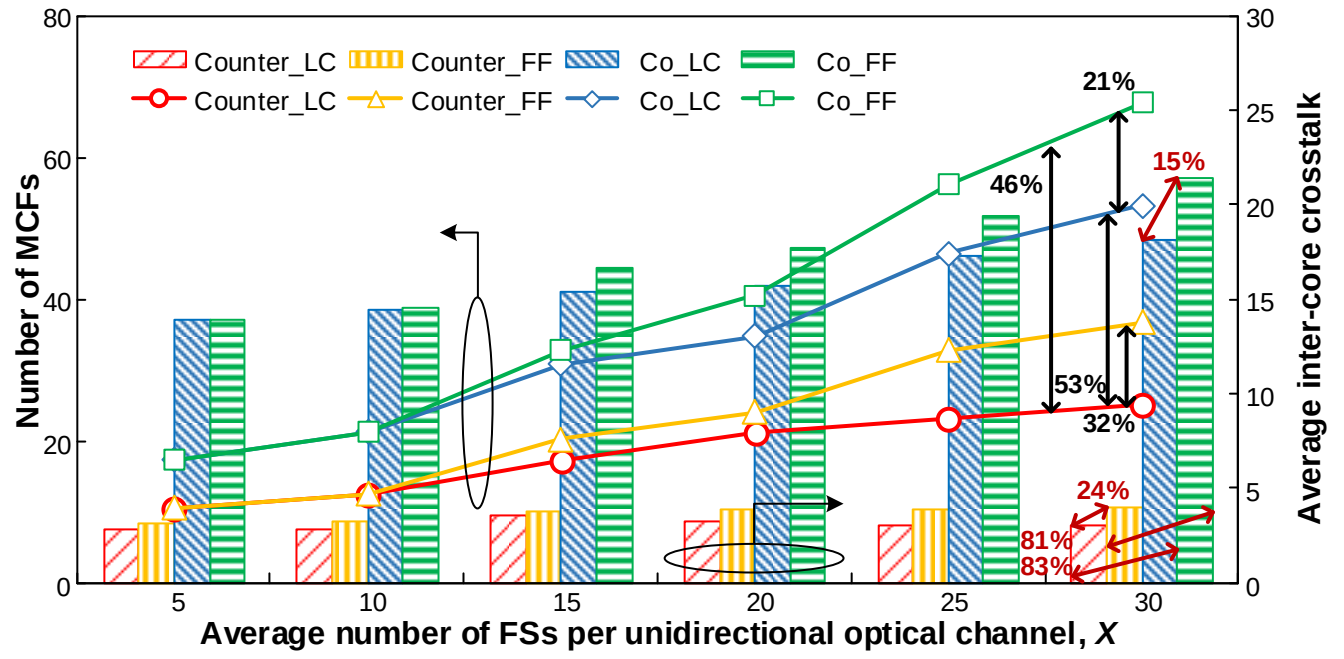

(b) 19-core MCF (NSFNET)

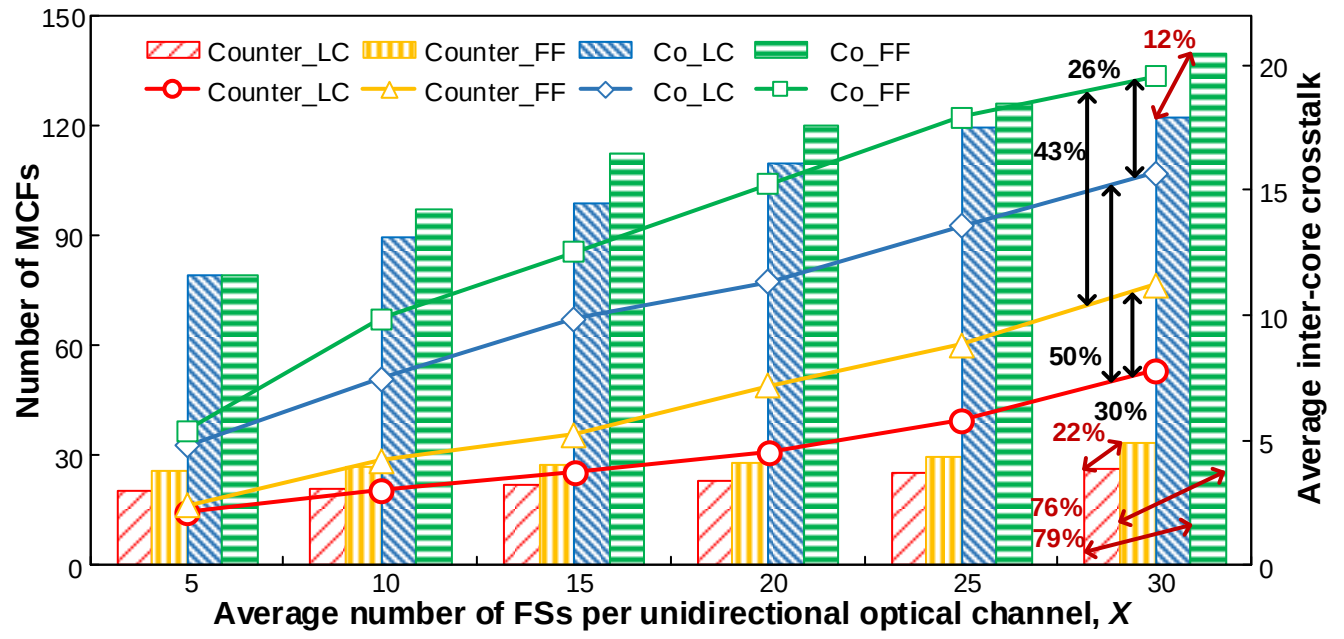

(c) 7-core MCF (COST239)

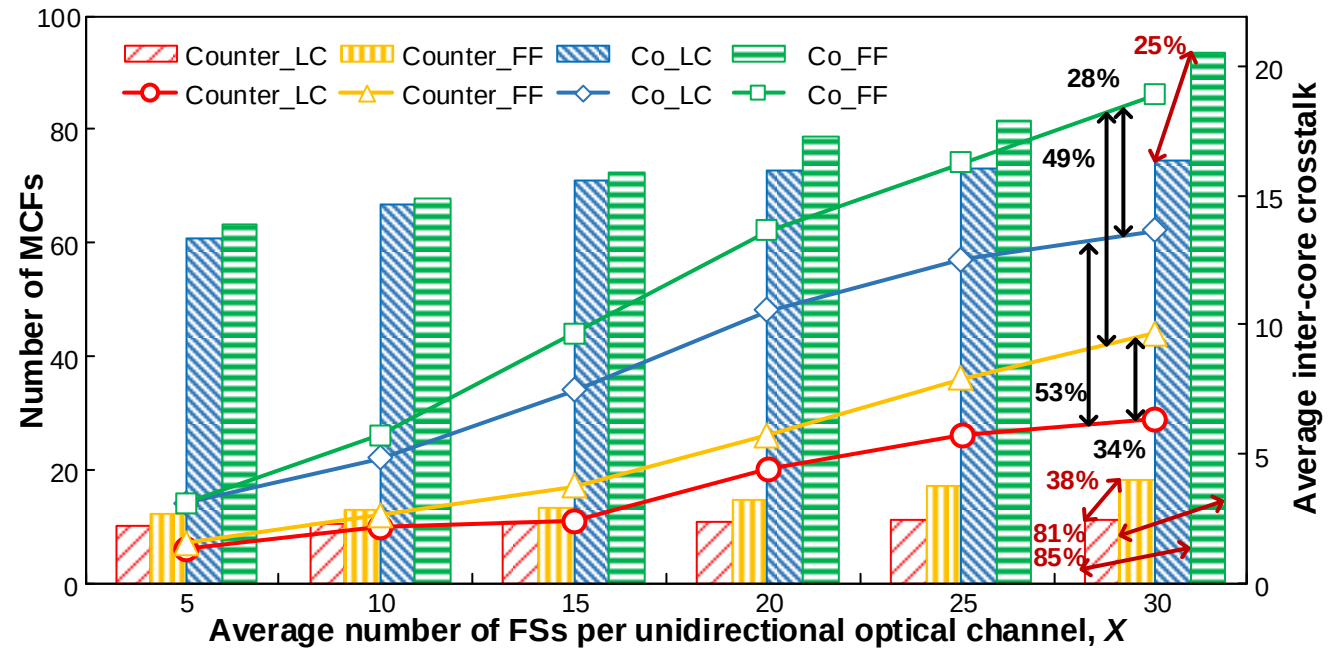

(d) 19-core MCF (COST239)

Fig. 11. Performance comparison in terms of number of MCFs used and inter-core crosstalk.

We also made similar performance comparisons for the other two test networks in Fig. 11. However, only the results of the heuristic algorithms are provided, because of the high computational complexity of the ILP model. Figs. 11(a) and (b) show the results of the NSFNET network for the 7-core and 19-core MCFs, respectively. Comparing the counter and co-propagation cases, we see that the former significantly reduces the number of MCFs used by up to 41% and 47% respectively for the FF and LC strategies in a 7-core MCF network and the reductions of 46% and 53% are even higher for a 19-core MCF network, respectively. In addition, the LC strategy outperforms the FF strategy by up to 19% and 27% respectively for counter and co-propagation situations in a 7-core MCF network and by 21% and 32% in a 19-core MCF network, respectively.

For the inter-core crosstalk, we find that the counter-propagation scheme can reduce the inter-core crosstalk by up to 74% and 77% respectively for FF and LC strategies in a 7-core MCF network. These values are 81% and 83% in a 19-core MCF network, respectively. In addition, we notice that an MCF network with more fiber cores, i.e., 19 vs. 7 cores, seems to show higher gains in reducing both the number of MCFs used and the inter-core crosstalk when the counter-propagation core assignment scheme is employed. This is reasonable since the 19-core MCFs network has more freedom in finding a core to serve a lightpath request and therefore has more chances in optimizing the core assignment.

We have similar observations for the results of the COST239 network. That is, the counter-propagation scheme with the LC strategy can achieve the best performance in terms of the number of MCFs used and inter-core crosstalk. These, once again, show the efficiency of the counter-propagation core assignment scheme.

### *B. Performance under Different Bidirectional Traffic Demand Asymmetric Ratios (ARs)*

We also evaluate how the counter-propagation core assignment mode can improve capacity utilization and inter-core crosstalk with different levels of bidirectional traffic demand asymmetry. In Fig. 12, we show how the number of MCFs used changes with an increasing asymmetric ratio (AR) in the n6s8 network. Here, to ensure a controllable AR between each node pair, we first generate a list of total FSs required by the bidirectional requests between each node pair which are uniformly distributed within the range of [5, 2$X$-5] FSs, where $X$ is the average number of FSs required and we set $X$=8 for this simulation study. We then divide each generated capacity requirement into two parts based on a ratio of 1:AR and the larger one is assigned to the unidirectional lightpath demand whose source node index is larger than that of the destination node for each node pair.

We see that with an increasing AR, the number of MCFs used under the counter-propagation mode does not change much. This is because the counter-propagation mode assigns different number of fiber cores in a common MCF according to the actual traffic demands in the two directions. This therefore avoids capacity and core wastage. Consequently, for a constant sum of bidirectional bandwidth requirement, it can be expected that the number of MCFs used will not change much for different ARs. In contrast, the co-propagation mode assumes that on each link, MCFs are deployed in pairs with each pair propagating optical signals in the opposite directions.

Moreover, the number of MCFs used is decided by the larger of the bandwidth requirements in the two directions. This therefore leads to much more MCFs being used in the co-propagation scheme when the AR increases. Specifically, there are 75% and 73% more MCFs used compared to the counter-propagation scheme respectively under the FF and LC strategies. Comparing the LC and FF strategies, we see that the former outperforms the latter by more than 15% because it selects a combination of route and SW with the fewest newly added MCFs and the least inter-core crosstalk. In addition, by comparing the results of the proposed spectrum and core assignment heuristic algorithm and the ILP model, we see the efficiency of the heuristic algorithm under the LC strategy is high and it performs close to the ILP model.

We have similar observations for the performance of inter-core crosstalk. With an increasing AR, the increase of inter-core crosstalk suffered by the counter-propagation mode is minor. In contrast, such an increase in the co-propagation scheme is much more significant. Specifically, the counter-propagation mode can reduce up to 80% inter-core crosstalk compared to the co-propagation mode for both the FF and LC strategies. In addition, the proposed spectrum and core assignment heuristic algorithm under the counter-propagation mode is very efficient as it shows an inter-core crosstalk close to that of the ILP model.

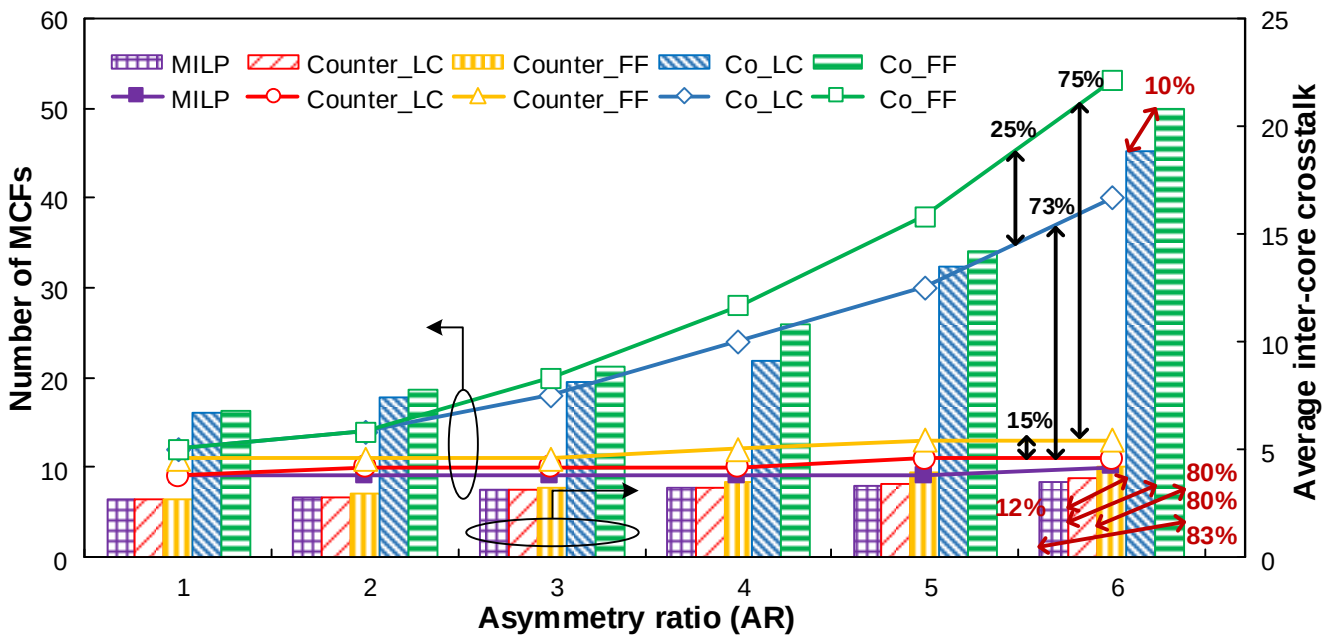

Fig. 12. Impact of traffic demand bidirectional asymmetry ratio.

We have conducted similar simulation studies for the other two larger test networks; these results are shown in Fig. 13. Here again due to the high computational complexity of solving the ILP model for these two larger networks, we only provide the results of the heuristic algorithm. We employed the same strategy as before in the n6s8 network, to assign asymmetric bidirectional capacity requirements for each node pair. The only difference is that we set a larger average number of FSs required by setting $X$=20. Figs. 13(a) and (b) show the results of the NSFNET network with 7 and 19 cores in each MCF, respectively. We see that for the 7-core scenario, the counter-propagation scheme is efficient to require almost constant numbers of MCFs with an increasing AR. Moreover, the counter-propagation mode can significantly reduce the number of MCFs used, by up to 63% and 58%, respectively, for the FF and LC strategies, compared to the co-propagation scheme. These values are 76% and 75% for the 19-core scenario, respectively. These therefore verify the efficiency of the counter-propagation scheme in avoiding capacity wastage due to bidirectional capacity asymmetry. In addition, comparing the LC and FF strategies, it is found that the former outperforms the latter by up to 25% and 15% respectively for the counter and co-propagation situations in a 7-core MCF network. These values are 25% and 19% in a 19-core MCF network, respectively. These confirm the efficiency of the LC strategy under different levels of ARs.

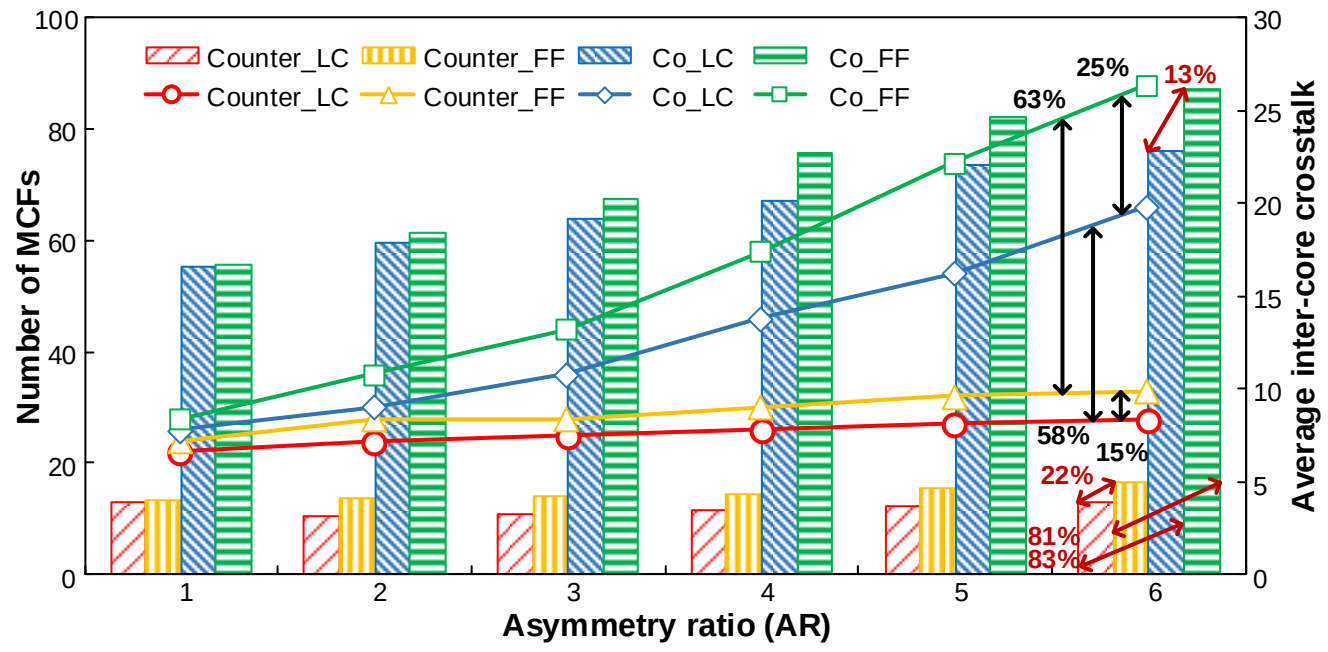

(a) 7-core MCF (NSFNET)

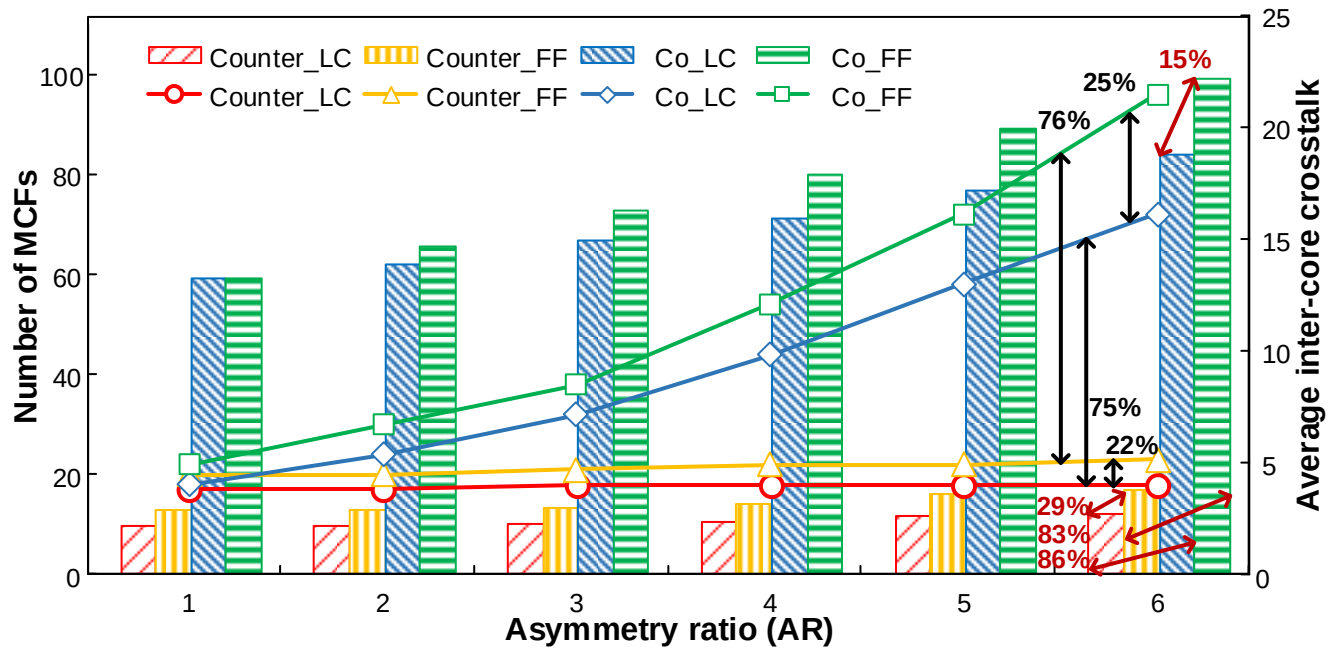

(b) 19-core MCF (NSFNET)

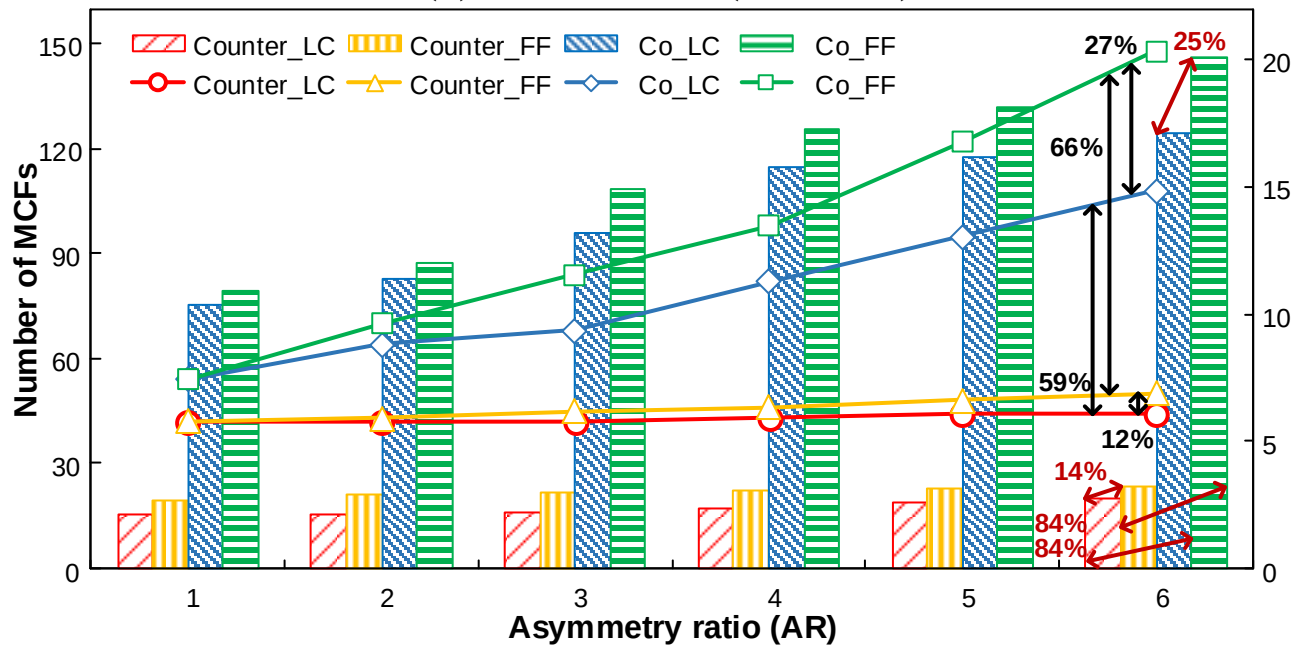

(c) 7-core MCF (COST239)

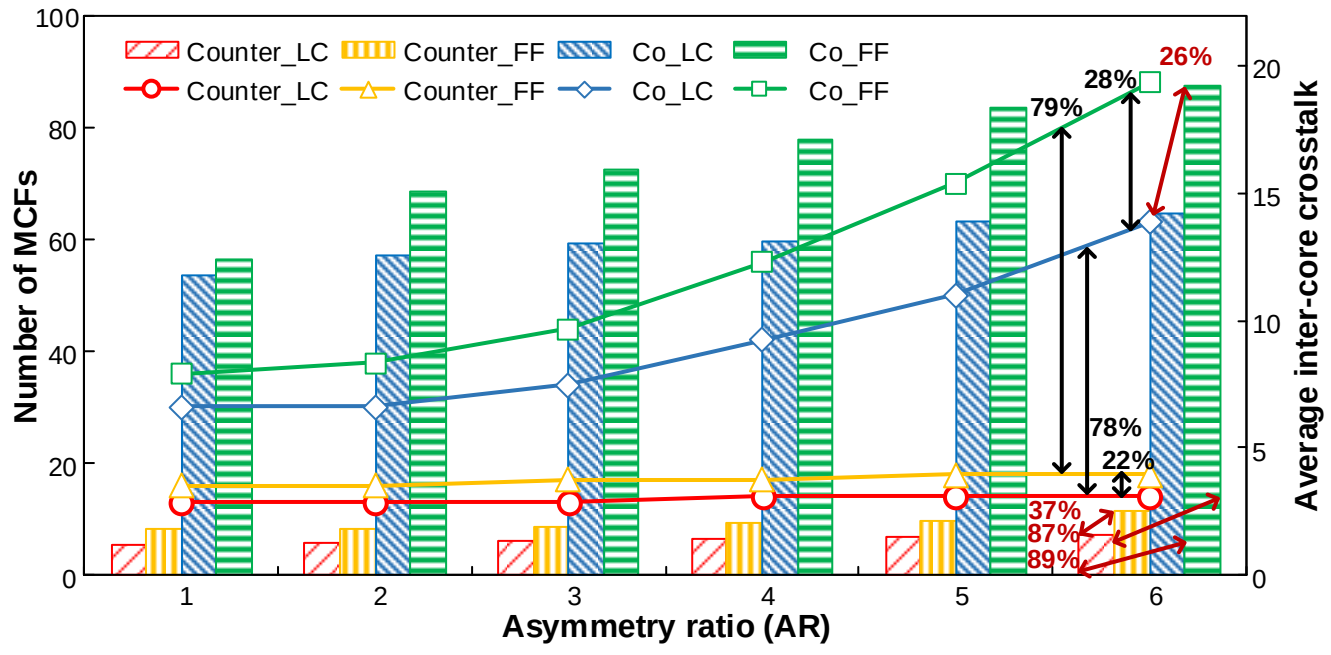

(d) 19-core MCF (COST239)

Fig. 13. Impact of traffic asymmetry ratio.

In addition, for the performance of inter-core crosstalk, the counter-propagation mode demonstrates a much better performance than the co-propagation mode. Specifically, the counter-propagation mode can reduce inter-core crosstalk by up to 81% and 83% respectively for the FF and LC strategies in a 7-core MCF network. These values are 83% and 86% in a 19-core MCF network. Similarly, the LC strategy significantly outperforms the FF strategy in reducing inter-core crosstalk for both the 7-core and 19-core networks.

Similar results are shown in Figs. 13(c) and (d) for the COST239 network. We reach the same conclusions as before, that the counter-propagation mode can achieve better performance, in terms of both the number of MCFs used and the inter-core crosstalk, than the co-propagation mode. In addition, the LC strategy can achieve much better performance than the FF strategy because the former explores all the possible spectrum and core assignment choices to select the one that is the most efficient.

### C. *Layouts of Fiber Cores*

In this section, we compare the layouts of fiber cores in each MCF by the different schemes based on the counter-propagation mode. We first compare the layouts corresponding to the results in Fig. 12 and find that the cores of two directions and the layout in each MCF of the ILP model and the most efficient heuristic scheme are almost the same. In addition, we also visually show the fiber core layouts of link N5-N9 in the NSFNET network based on the simulation studies of Figs. 13(a) and (b) (see Fig. 14). We can see that when AR=1, which correspond to the case of symmetric bidirectional traffic demand, the usage of fiber cores in the two opposite directions are very close, i.e., 3 vs. 4 fiber cores. These counter-propagating fiber cores are arranged in an interleaving manner, which ensures the lowest inter-core crosstalk. In contrast, for a larger bidirectional AR, i.e., AR=6, we see that there are different numbers of fiber cores in the two propagation directions, which is 2 vs. 5. The fiber core layout still follows the interleaving manner for minimum inter-core crosstalk. Note that for the scenario of a higher AR (see Fig. 13(b)), the "IN" fiber cores on the left-hand side would lead to strong inter-core crosstalk as they are transmitting signals in the same direction. However, such a layout is reasonable since in our optimization, the major objective is to minimize the total number of MCFs used. When assigning spectra and cores for lightpaths, we always try to fill up or fully use the cores in each MCFs at the first priority even though there can be some inter-core crosstalk. However, the two "OUT" cores are properly separated, which can minimize inter-crosstalk.

For the 19-core scenario, we also have similar fiber-core layouts as shown in Fig. 14. Under the symmetric case, i.e., AR=1, the fiber cores are well separated from each other in both directions to minimize inter-core crosstalk. Moreover, for the asymmetric case, i.e., AR=6, though there is inter-core crosstalk between "IN" cores, whose number is larger than that of "OUT" cores, the "OUT" cores are well separated to have the smallest inter-core crosstalk. All these fiber-core layouts visually demonstrate the effectiveness of the proposed spectrum and core assignment algorithm based on the counter-propagation mode for minimizing inter-core crosstalk.

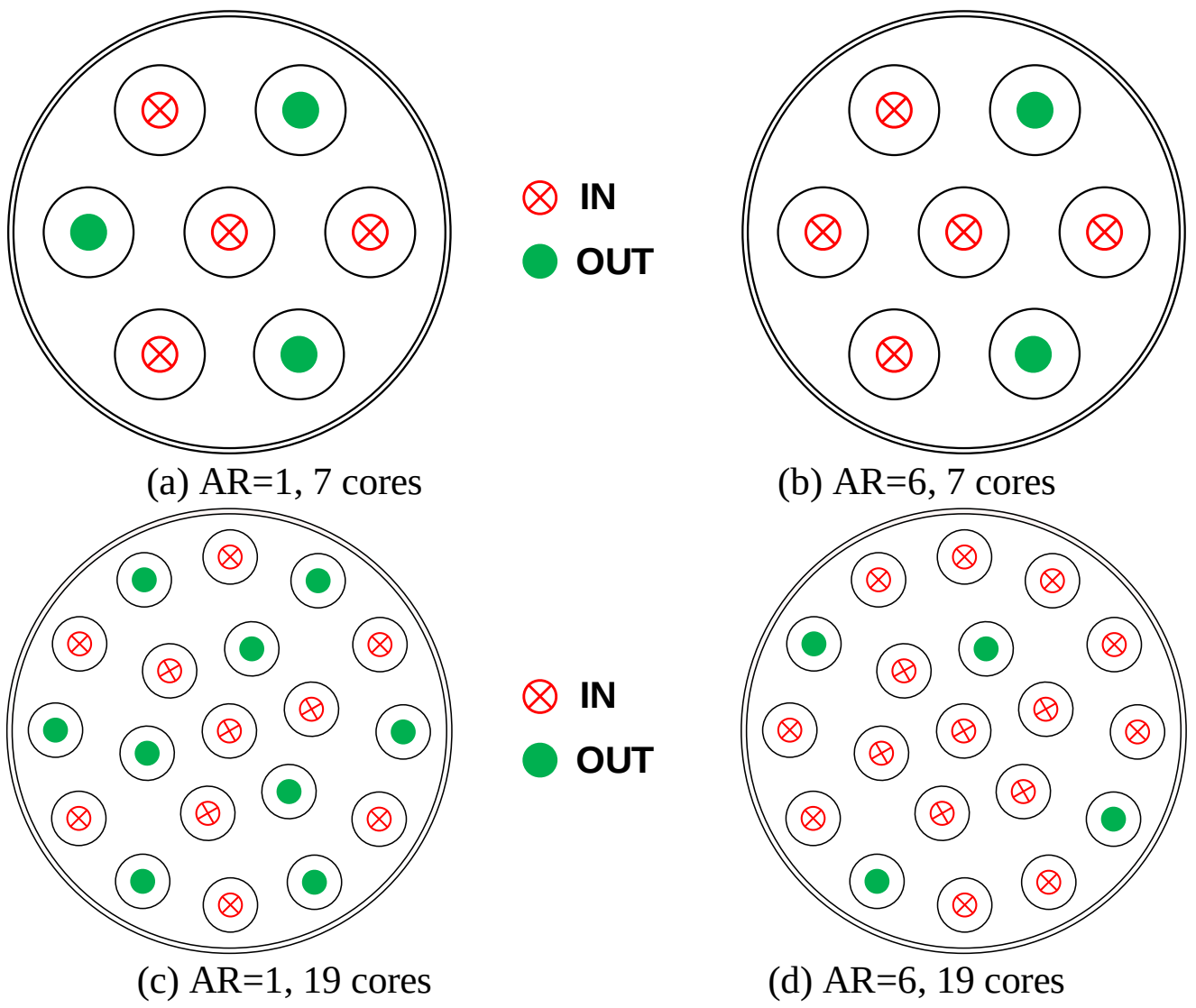

Fig. 14. Fiber core layouts in MCFs (link N5-N9, NSFNET).

## VII. Conclusions

We proposed the counter-propagation mode of fiber cores in an MCF core when assigning spectra and cores for lightpath establishment. This new mode can greatly reduce inter-core crosstalk and capacity wastage due to bidirectional traffic demand asymmetry. We evaluated the effectiveness of the proposed scheme in the context of the routing, spectrum, and core assignment (RSCA) problem for an MCF optical network. An ILP model was developed for the problem and an efficient RSCA algorithm was further proposed to assign spectra and cores in each fiber link in an interleaving manner to minimize the number of MCFs used and the inter-core crosstalk. The simulation studies show that the proposed counter-propagation mode is effective as it significantly reduces the number MCFs used as well as the inter-core crosstalk, compared to the conventional MCF optical network with the fiber-core co-propagation mode. In addition, the proposed spectrum and core assignment algorithm with the LC strategy is efficient enough to perform very close to the ILP model and does much better than the FF strategy. Finally, under different levels of traffic demand ARs, the counter-propagation mode also demonstrates much better performance in terms of the MCFs used and inter-core crosstalk. An almost constant number of MCFs was required and a minor change in inter-core crosstalk was demonstrated for the counter-propagation mode to carry traffic demands with different bidirectional ARs. All these results therefore verify the effectiveness of the proposed counter-propagation core assignment mode for reducing both the inter-core crosstalk and the number of MCFs used, in an MCF optical network.